\documentclass[12pt,a4paper,final]{iopart}
\usepackage{iopams}  
\usepackage{graphicx}
\begin{document}

\title{Characteristics of a magneto-optical trap of molecules}

\author{H J Williams, S Truppe, M Hambach, L Caldwell, N J Fitch, E A Hinds, B E Sauer and M R Tarbutt}
\address{Centre for Cold Matter, Blackett Laboratory, Imperial College London, Prince Consort Road, London SW7 2AZ UK}
\eads{hannah.williams10@imperial.ac.uk, m.tarbutt@imperial.ac.uk}

\begin{abstract}
We present the properties of a magneto-optical trap (MOT) of CaF molecules. We study the process of loading the MOT from a decelerated buffer-gas-cooled beam, and how best to slow this molecular beam in order to capture the most molecules. We determine how the number of molecules, the photon scattering rate, the oscillation frequency, damping constant, temperature, cloud size and lifetime depend on the key parameters of the MOT, especially the intensity and detuning of the main cooling laser. We compare our results to analytical and numerical models, to the properties of standard atomic MOTs, and to MOTs of SrF molecules. We load up to $2 \times 10^{4}$ molecules, and measure a maximum scattering rate of $2.5 \times 10^{6}$~s$^{-1}$ per molecule, a maximum oscillation frequency of 100~Hz, a maximum damping constant of 500~s$^{-1}$, and a minimum MOT rms radius of 1.5~mm. A minimum temperature of 730~$\mu$K is obtained by ramping down the laser intensity to low values. The lifetime, typically about 100~ms, is consistent with a leak out of the cooling cycle with a branching ratio of about $6 \times 10^{-6}$. The MOT has a capture velocity of about 11~m/s.

\end{abstract}

\vspace{2pc}

%\maketitle
%\newpage

\section{Introduction}

The magneto-optical trap (MOT)~\cite{Raab1987} is at the heart of a vast range of scientific and technological applications that use ultracold atoms. In a MOT, pairs of counter-propagating laser beams cross at the zero of a magnetic quadrupole field, subjecting atoms to a velocity-dependent force, which cools them, and a position force, which traps them. Recently, there has been a great effort to extend the method to molecules, motivated by many new applications~\cite{Carr2009} that include quantum simulation and information processing, the study of collisions and ultracold chemistry, and tests of fundamental physics. Laser cooling has been applied to a few species of diatomic molecules~\cite{Shuman2010, Hummon2013, Zhelyazkova2014}, and recently even to a triatomic molecule~\cite{Kozyryev2017}. Two-dimensional magneto-optical compression of a YO beam was demonstrated~\cite{Hummon2013}, followed by the first three-dimensional molecule MOT~\cite{Barry2014}, which used SrF. 

Laser cooling and magneto-optical trapping are more difficult for molecules than for atoms. Several vibrational branches have to be addressed, each requiring a separate laser. For molecules with electronic and nuclear spin, spin-rotation and hyperfine interactions further increase the number of levels involved. To avoid decay to multiple rotational states it is necessary to drive transitions with $F \ge F'$~\cite{Stuhl2008}, where $F$ and $F'$ are the angular momenta of the ground and excited states. Consequently,  the number of ground states typically exceeds the number of excited states, and this reduces the scattering rate and increases the  saturation intensity~\cite{Tarbutt2013}. Furthermore, for such transitions, there are dark states present amongst the ground-state sub-levels. This can be especially problematic in a MOT because optical pumping into dark states diminishes the trapping force~\cite{Tarbutt2015}. In the rf MOT~\cite{Norrgard2016, Steinecker2016} this problem is avoided by synchronously modulating the laser polarization and magnetic field direction at frequencies similar to the scattering rate, typically around 1~MHz. In a dc MOT, where there is no such modulation, one can either rely on the motion of the molecules through the spatially-varying polarization of the light field to de-stabilize the dark states, or make use of the dual-frequency mechanism~\cite{Tarbutt2015b}. Here, one (or more) of the MOT transitions is driven by two frequency components with opposite circular polarization, one red-detuned and the other blue-detuned. Both types of MOT have been investigated for SrF. For this molecule, the dc MOT exhibited a short lifetime and sub-millikelvin temperatures could not be reached~\cite{Barry2014, McCarron2015}. In the rf MOT the lifetime was longer and it was found that the temperature could be reduced to 250~$\mu$K by ramping down the intensity of the MOT light~\cite{Norrgard2016, Steinecker2016}. Recently, we demonstrated a dc MOT of CaF molecules, and we showed how to cool the molecules to 50~$\mu$K by first ramping down the intensity and then transferring to an optical molasses where sub-Doppler cooling processes are effective~\cite{Truppe2017b}. An rf MOT of CaF has also now been demonstrated~\cite{Anderegg2017}. 

Since molecular MOTs display some important differences to atomic MOTs, and since they are still new, a thorough characterisation of their properties is called for. In this paper, we study our dc MOT of CaF in detail. The MOT is loaded from a beam of molecules decelerated to low velocity by frequency-chirped counter-propagating laser light. We investigate how the parameters of this deceleration step influence the number of molecules loaded into the MOT, and how to maximize this number. Then, we study how the molecule number, scattering rate, trap oscillation frequency, damping constant, temperature, cloud size and lifetime each depend on the intensity and detuning of the MOT light. We also measure how the number of molecules and the cloud size depend on the applied magnetic field gradient, and we measure the capture velocity of the MOT. We compare our results to simple analytical models and find that most of the dependencies we observe are described adequately by these models. Thus, despite their greater complexity, many properties of molecular MOTs can be understood using similar models as for atomic MOTs, with only minor modifications. We also compare our measurements to the results of numerical models based on multi-level rate equations~\cite{Tarbutt2015} and find good agreement for most, but not all, of the MOT properties.

\section{Methods}

\begin{figure}[tb]
	\centering
	\includegraphics[scale=0.8]{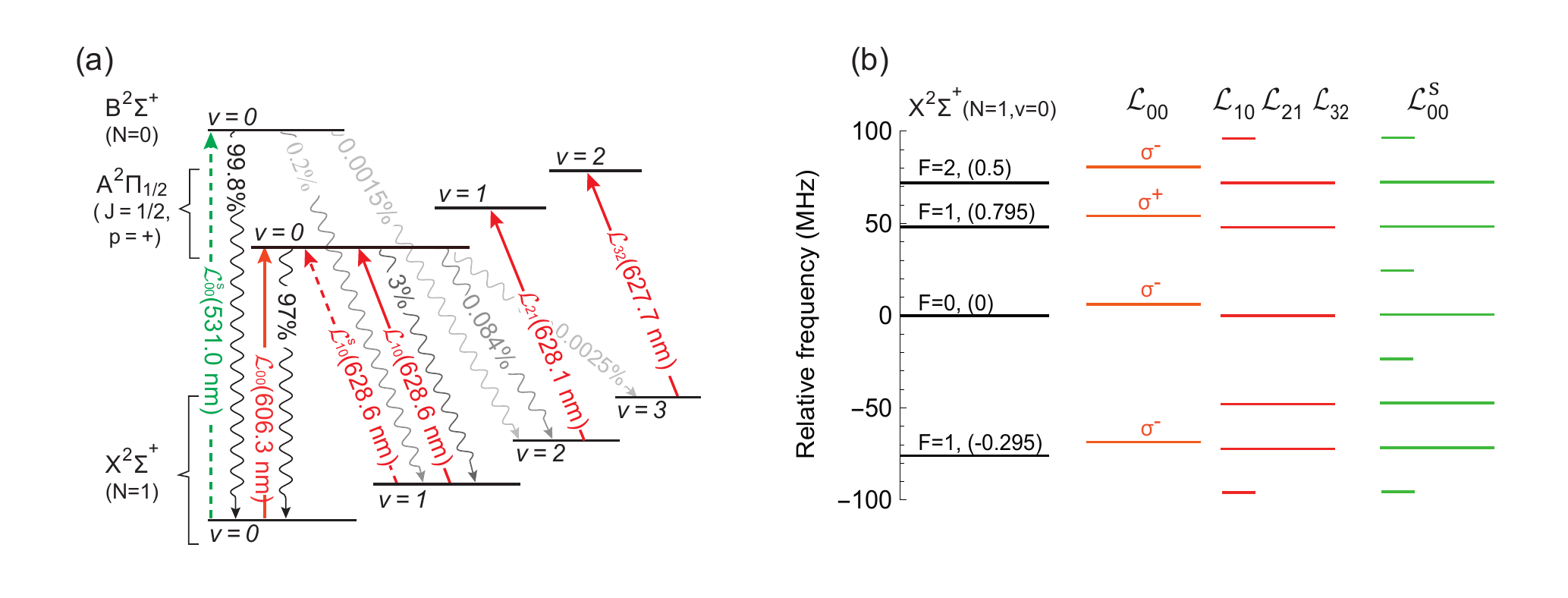}
	\caption{(a) Transitions driven in the experiment, with wavelengths and branching ratios. Dashed lines are the transitions driven by the slowing lasers, while solid lines are those driven by the MOT lasers. (b) Hyperfine levels of the ground state and sideband structure of the MOT lasers, with line length indicating the relative intensity of each sideband. The polarisation of each component of ${\cal L}_{00}$ is also shown. $v$, $N$, $J$, $F$ and $p$ are the quantum numbers of vibration, rotational angular momentum, total angular momentum excluding nuclear spin, total angular momentum, and parity. The magnetic $g$-factor of each hyperfine component is given in brackets.}
	\label{fig:structure}
\end{figure}

The setup used for this work is the same as described in \cite{Truppe2017b}. Here we give a more detailed description of the setup and our methods. The MOT is loaded from a buffer gas source of CaF molecules that are decelerated to low velocity using frequency-chirped counter-propagating light. The setup uses five lasers, which are detailed in table \ref{tab:lasers}. The transitions driven by these lasers are illustrated in figure \ref{fig:structure}(a) which also specifies the molecular notation we use. The ground electronic state of CaF is $X^2 \Sigma ^+$, the first electronically excited state is $A^2 \Pi$ whose decay rate is $2\pi\times 8.3$~MHz, and the second excited state is $B^2 \Sigma ^+$ whose decay rate is $2\pi\times 6.3$~MHz. The main slowing laser drives the $B^2 \Sigma ^+ (v=0,N=0) \leftarrow X^2 \Sigma ^+ (v=0,N=1)$ transition and is denoted ${\cal L}^{\rm s}_{00}$. A repump laser, ${\cal L}^{\rm s}_{10}$, drives the  $A^2 \Pi_{1/2} (v=0, J=1/2, p=+) \leftarrow X^2 \Sigma ^+ (v=1,N=1)$ transition, to recover the population that leaks into the $v=1$ state. The MOT uses four lasers which drive $A^2 \Pi_{1/2} (v=j,J=1/2,p=+) \leftarrow X^2 \Sigma ^+ (v=i,N=1)$ transitions and are denoted ${\cal L}_{ij}$. Although the vibrational branching ratios from the B state are more favourable than for the A state, the $B^2 \Sigma ^+ (v=0,N=0)$ state has a hyperfine splitting of about 20~MHz which might be problematic for a MOT, being neither large nor small compared to the linewidth~\cite{Tarbutt2015b}. The hyperfine splitting in the $A^2 \Pi_{1/2} (v=0,J=1/2,p=+)$ is unresolved, which is why we preferred to use this state for the MOT. Each of the lower levels, $X^2 \Sigma^+(v,N=1)$, is split into two components by the spin-rotation interaction, and these are each split in two again by the hyperfine interaction, giving four components\footnote{For simplicity, we refer to these as the hyperfine components.} with total angular momentum $F=1,0,1,2$. Figure \ref{fig:structure}(b) shows the hyperfine intervals for the $X^2 \Sigma ^+ (v=0,N=1)$ state. The hyperfine intervals for the other vibrational states are similar. Figure \ref{fig:structure}(b) also shows the sideband structure applied to each of the MOT lasers to ensure that all hyperfine components of each transition are driven (see later).

\fulltable{\label{tab:lasers} Lasers used in the experiment, the transitions they drive, and their frequencies and powers.}
\br

Laser & Transition$^{\rm a}$ & Role & Frequency (GHz)$^{\rm b}$ & Power (mW)$^{\rm c}$ \\
\mr
${\cal L}^{\rm s}_{00}$ & $B(0) \leftarrow X(0) $& Slowing & 564581.3 & 120 \\
${\cal L}^{\rm s}_{10}$ & $A(0) \leftarrow X(1) $& Slowing repump & 476958.0 & 130\\
${\cal L}_{00}$ & $A(0) \leftarrow X(0)$ & MOT & 494431.3 & 80\\
${\cal L}_{10}$ & $A(0) \leftarrow X(1)$ & MOT $v=1$ repump$^{\rm d}$ & 476958.0 & 100\\
${\cal L}_{21}$ & $A(1) \leftarrow X(2)$ & MOT $v=2$ repump & 477298.7 & 10\\
${\cal L}_{32}$ & $A(2) \leftarrow X(3)$ & MOT $v=3$ repump & 477627.6 & 0.5\\
\br
\end{tabular*}
$^{\rm a}$$X(i): X^2 \Sigma ^+ (v\!=\!i, N\!=\!1)$; $A(i): A^2 \Pi_{\frac{1}{2}} (v\!=\!i,J\!=\!\frac{1}{2},p\!=\!+)$; $B(i): B^2 \Sigma ^+ (v\!=\!i, N\!=\!0)$.\\
$^{\rm b}$Accurate to 600~MHz.\\
$^{\rm c}$Full power in a single beam.\\
$^{\rm d}$This light is derived from the same laser as ${\cal L}^{\rm s}_{10}$.
\end{table}

The source of CaF molecules is a cryogenic buffer gas source, described briefly here and in detail in reference~\cite{Truppe2017c}. A calcium target inside a 4~K copper cell is ablated at $t=0$ by a pulse of light from a Nd:YAG laser ($5$~mJ energy, $4$~ns duration, $1064$~nm wavelength, 2~Hz repetition rate). The liberated Ca enters a stream of 4~K helium gas flowing through the cell with a flow rate of 0.5~sccm. Sulphur hexafluoride (SF$_6$) enters the cell from a room temperature capillary at a flow rate of $0.01$~sccm. The Ca and SF$_6$ react to create CaF molecules which are cooled by collisions with the He and leave the cell through a $3.5$~mm aperture. The resulting pulse contains around $1.9 \times 10^{11}$ CaF molecules per steradian in the $X^2 \Sigma ^+ (v=0,N=1)$ state, with an average forward velocity of $150$~m/s and a duration of $280$~$\mu$s measured $2.5$~cm downstream of the exit aperture. The beam then passes through an $8$~mm diameter skimmer, $15$~cm downstream, which separates the source chamber from the slowing chamber where the molecules are decelerated to low speed. Finally, they enter the MOT chamber via a $20$~mm diameter, $200$~mm long differential pumping tube, and are captured in the MOT 120~cm from the exit of the source. The pressures in the source, slowing and MOT chambers are $2\times10^{-7}$, $6\times10^{-8}$ and $2\times10^{-9}$~mbar respectively. The MOT and slowing chambers are connected via a bellows for vibration isolation.

\begin{figure}
\centering
\includegraphics[scale=0.7]{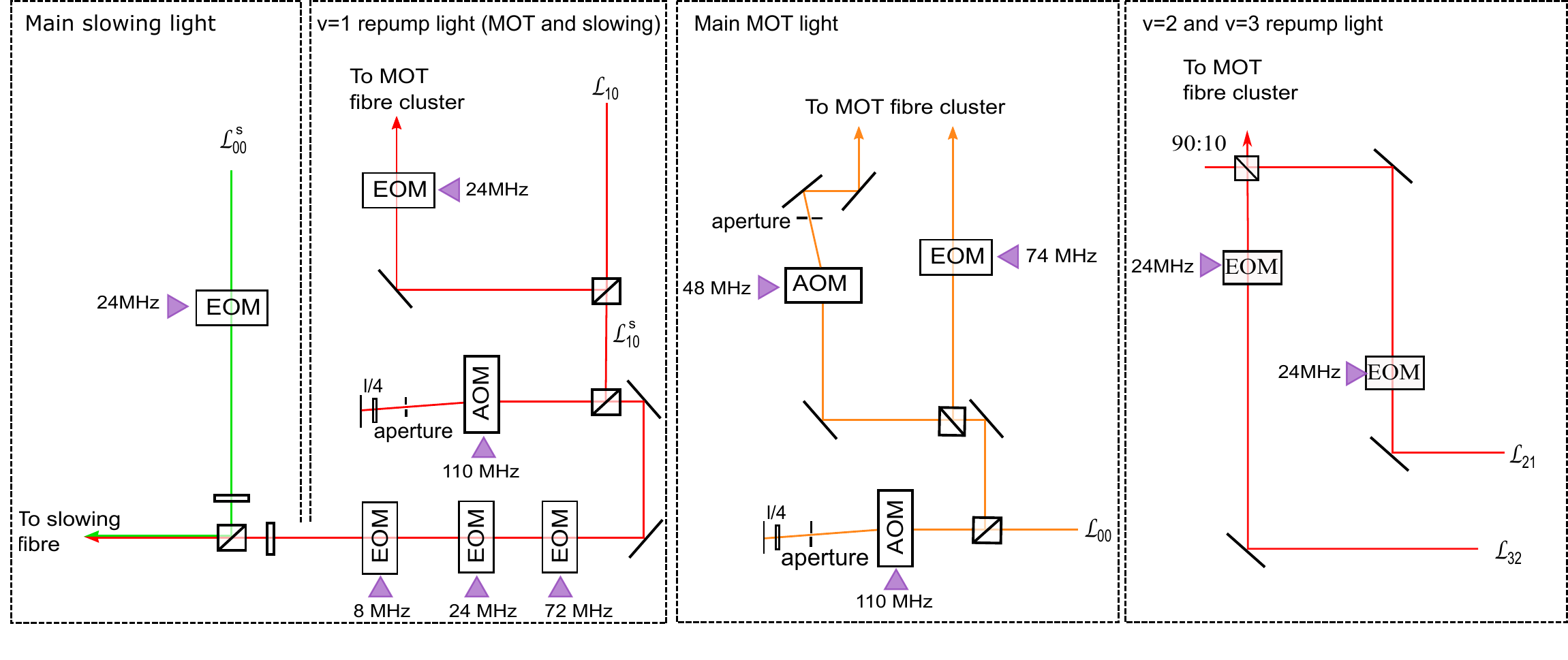}
\caption{Schematic showing how sidebands are applied to the laser light to address the hyperfine structure, and how the beams are combined.}
\label{fig:lasers}
\end{figure}

The molecules are decelerated using the frequency-chirped slowing technique described in reference \cite{Truppe2017}. The slowing light counter-propagates to the molecular beam and consists of ${\cal L}^{\rm s}_{00}$ and ${\cal L}^{\rm s}_{10}$ combined into a single beam with a Gaussian intensity profile whose $1/e^2$ radius is $9$~mm at the MOT converging to $1.5$~mm at the source. It is linearly polarized at $45^\circ$ to the direction of a uniform 0.5~mT magnetic field applied throughout the slowing region that de-stabilizes dark Zeeman sub-levels. Figure \ref{fig:lasers} shows the setup used to apply rf sidebands to the laser light and then combine the beams. The spectrum of ${\cal L}^{\rm s}_{00}$ is shown in figure \ref{fig:structure} and is generated by driving an electro-optic modulator (EOM) at 24~MHz with a phase modulation index of approximately 3.1. The initial detuning of ${\cal L}^{\rm s}_{00}$ is set to $-270$~MHz so that molecules travelling at $145$~m/s are Doppler shifted into resonance.  ${\cal L}^{\rm s}_{00}$ is turned on at $t=t_{{\rm on}}$ with the frequency held constant until $t=t_{{\rm chirp}}$. Then, the frequency is linearly chirped at rate $\alpha$ from $t=t_{{\rm chirp}}$ to  $t=t_{{\rm off}}$ when the light is turned off. This chirp, which is applied directly to the laser, ensures that the light remains resonant with the molecules as they slow down. The repump light, ${\cal L}^{\rm s}_{10}$, is derived from the same laser as ${\cal L}_{10}$ which has zero detuning. ${\cal L}^{\rm s}_{10}$ passes twice through a 110~MHz acousto-optic modulator (AOM) which is used to control its detuning, nominally -220~MHz. The frequency-shifted light then passes through three successive EOMs driven at $72$, $24$ and $8$~MHz giving a near-continuous spectrum of light with a width of about 360~MHz.\footnote{90\% of the power lies within this bandwidth.} The frequency shift and broadening ensure that all molecules interact strongly with the light irrespective of velocity or hyperfine state. ${\cal L}^{\rm s}_{10}$ is turned on at $t_{{\rm on}}$ and off at $t_{{\rm off}}$, and its centre frequency is constant throughout.

Figure \ref{fig:structure}(b) shows the frequency components of the main MOT light, ${\cal L}_{00}$, and the handedness of each component. The transition from $F=2$ is driven by two frequency components of opposite handedness, one red-detuned and the other blue-detuned. This implements the dual-frequency scheme described in \cite{Tarbutt2015b}, which avoids optical pumping into dark states and produces strong confinement. Because the separation between the $F=2$ and upper $F=1$ hyperfine levels is only 24~MHz, the laser component detuned to the red of the $F=1$ level also acts as the blue-detuned component for the $F=2$ level. The optical system used to generate the required light is shown in figure \ref{fig:lasers}. We first split ${\cal L}_{00}$ into two parts in the ratio $3:1$. The first part passes through a $74.5$~MHz EOM, so that the carrier addresses $F=0$ while the sidebands address the $F=2$, and lower $F=1$ levels. The second passes through a 48~MHz AOM to generate the sideband that addresses the upper $F=1$ level. The detuning and power of ${\cal L}_{00}$ are both varied throughout the experiments presented in this paper. They are controlled by a $110$~MHz AOM set up in a double-pass configuration. Each of the three MOT repumps passes through a $24$~MHz EOM to generate the sideband structure shown in figure \ref{fig:structure}(b). All three are set to zero detuning. The two ${\cal L}_{00}$ beams along with the three MOT repumps are combined using a fibre cluster\footnote{Schafter \& Kirchhoff} into a single fibre that delivers the light to the MOT. The output of the fibre is linearly polarized and the intensity profile has a $1/e^2$ radius of $8.1$~mm. 

The resonant frequency of each laser is given in table \ref{tab:lasers}. We find these by using each laser, with sidebands, as a transverse probe, and measuring the laser-induced fluorescence as a function of frequency. The largest fluorescence signal occurs when all hyperfine levels are addressed simultaneously. For all lasers other than ${\cal L}_{00}$ we refer to this as zero detuning. For ${\cal L}_{00}$ we find that there is a critical frequency where a MOT is produced in half of all shots, and we denote this as the zero of detuning for ${\cal L}_{00}$, $\Delta_{00} =0$. No MOT is formed when $\Delta_{00}>0$, and a stable MOT is formed for $\Delta_{00}<0$. This is a more sensitive and reproducible way of fixing the frequency than finding the maximum fluorescence, which occurs at $\Delta_{00}=2\pi\times 2(4)$~MHz.

\begin{figure}[t]
	\centering
	\includegraphics{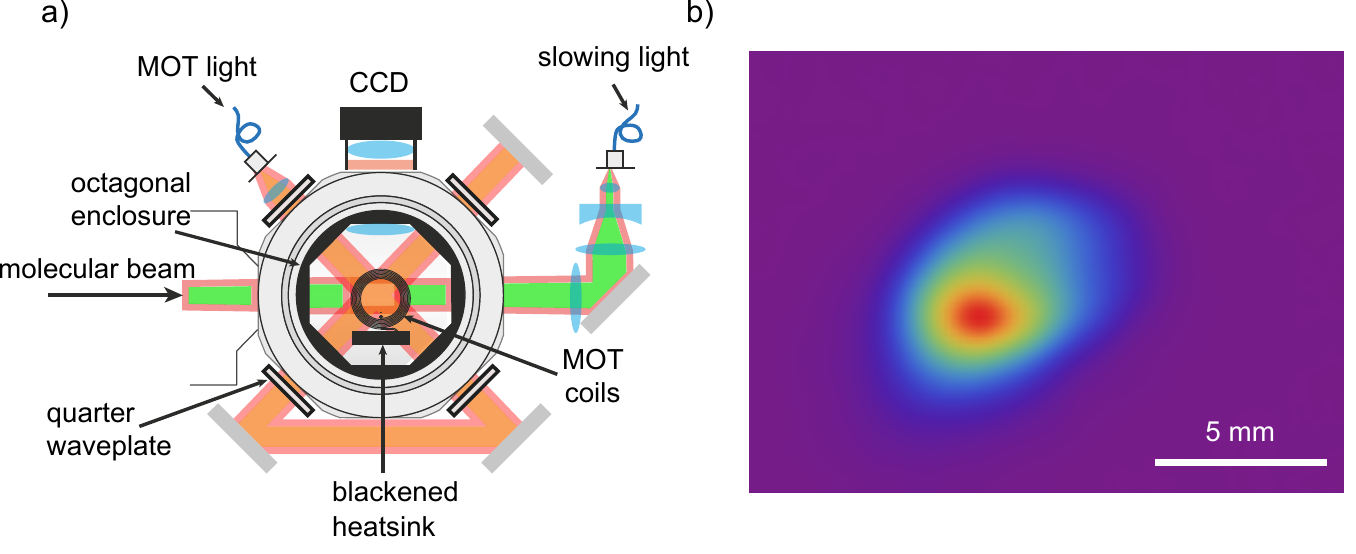}
	\caption{(a) Setup of the MOT showing the vacuum chamber, the horizontal MOT beams and the imaging setup. (b) Fluorescence image of a single MOT (no averaging) using an exposure time of 50~ms.
	}
	\label{fig:setUp}
\end{figure}

Figure \ref{fig:setUp} shows the MOT chamber and illustrates how the light is folded and retro-reflected to produce the six counter-propagating MOT beams. The optical layout follows reference \cite{Barry2014} -- at each input window a 617~nm quarter-wave plate produces the required circular polarization, and at each exit window another quarter-wave plate returns the polarization to linear, so that the light is linearly polarized at every mirror. The vertical beams have opposite handedness to the horizontal ones. All windows and waveplates have a broadband anti-reflection coating. The beam is slightly converging to compensate for the residual losses, so that each pass has approximately the same intensity. To ensure optimal overlap of the retro-reflected light the beam is recoupled back through the fibre. 

The magnetic quadrupole field of the MOT is produced by a pair of anti-Helmholtz coils, with an inner diameter of 30~mm, placed inside the MOT chamber as illustrated in figure~\ref{fig:setUp}(a). Each coil consists of two laser-cut copper spirals, with eight turns per spiral, mounted on either side of an AlN plate. The plates are connected to a copper block which is mounted to the chamber and acts as a heat sink. Three Helmholtz coil pairs mounted around the chamber, one for each axis, are used to cancel background fields. To take pictures of the MOT, we image its fluorescence. The imaging system is composed of two lenses, a 50.8~mm diameter, $60$~mm focal length lens inside the chamber which collimates the fluorescence, and a 40~mm diameter, 28.6~mm focal length lens outside the chamber, 100~mm from the first, which images the light onto either a photomultiplier tube (PMT) or a CCD camera. The imaging system has a measured magnification factor of $0.5$. From numerical ray tracing, transmission measurements and the specified quantum efficiency of the camera, we calculate a detection efficiency of $1.5(2)\%$.  We were careful to reduce background scatter from the MOT beams. The MOT coils, the AlN support plates, and an octagonal enclosure inside the chamber are all painted black\footnote{Alion MH2200.}. The copper heat sink is covered with light absorbing foil\footnote{Acktar Spectral Black.} which provides a dark background against which the MOT fluorescence is imaged. A bandpass filter placed between the two lenses transmits the dominant 606~nm fluorescence while blocking the background scatter from the 628~nm lasers. With this set up, and the MOT beams at full power, the background is $8.5\times 10^{4}$~photons/s/mm$^2$. This may be compared to the fluorescence collected from a single molecule which is $3.6\times 10^{4}$~photons/s. Figure \ref{fig:setUp}(b) shows an image of a MOT. From the number of photons collected, our measurement of the scattering rate (see section \ref{sec:scatteringRate}) and the measured detection efficiency, we deduce that there are $4.6 \times 10^{3}$ molecules in this MOT. When all parameters are optimized, this number is about $2\times10^4$. We typically sum together 50 images, giving a standard deviation in the number of molecules detected of 3\%.

\section{Loading the MOT}
\label{sec:MOTLoading}

\begin{figure}
\centering
\includegraphics[scale=0.8]{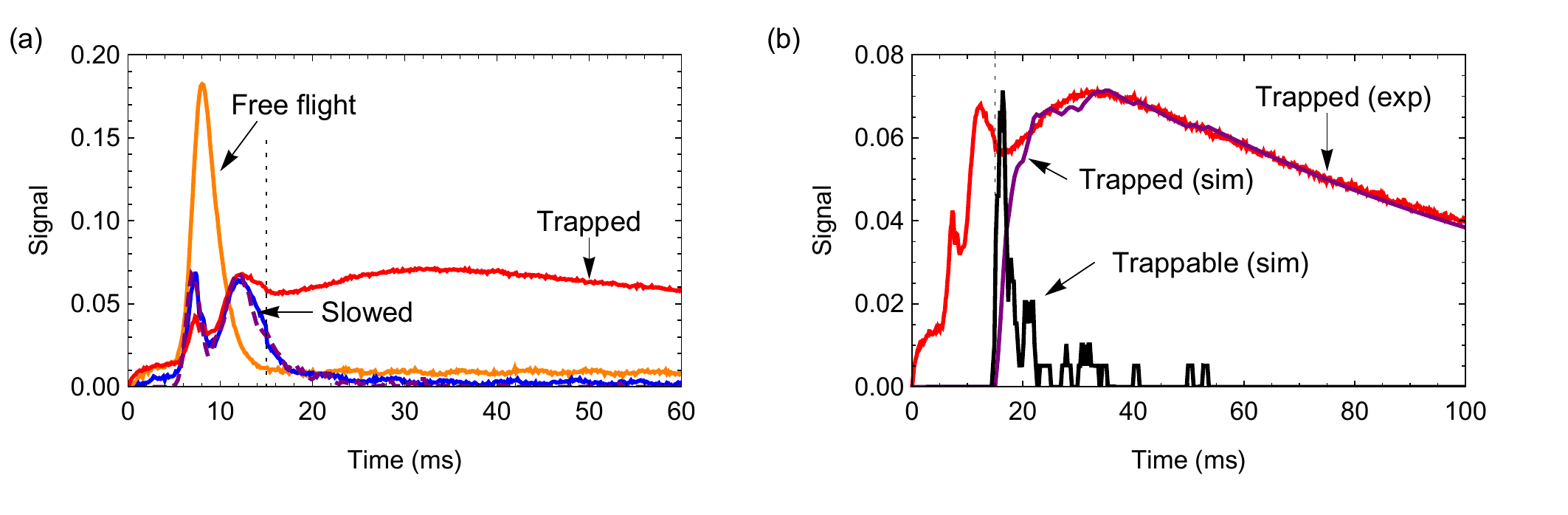}
\caption{(a) Fluorescence from the MOT region detected on a PMT, as a function of time.  Orange: free flight (experiment); Blue: slowed (experiment); Dashed purple: slowed (simulation); Red: trapped molecules (experiment). The slowing chirp ends at $t_{\rm off}=15$~ms (vertical dotted line), giving $\alpha (t_{\rm off}-t_{\rm chirp}) = 285$~MHz. The powers of ${\cal L}_{00}^{\rm s}$ and ${\cal L}_{10}^{\rm s}$ are $120$~mW and $130$~mW respectively. The orange and blue curves are recorded with a single beam of the MOT light and no magnetic field, whereas the red curve is recorded with all six beams and the magnetic field on. The red curve has been scaled down by a factor of 5 to account for the increased fluorescence resulting from of all six beams. (b) Comparison of experiment and simulation for MOT loading. Red: experimental data from (a). Black: simulation results showing the arrival-time distribution of molecules that arrive later than 15~ms, pass through a $1$~cm diameter disk in the plane of the MOT, and have velocity below $10$~m/s. Purple: integration of the black signal multiplied by an exponential decay with a time constant of $95$~ms, and scaled to best match the experimental data.}
\label{fig:MOTLoading}
\end{figure}

We will see in section \ref{sec:captureVelocity} that the capture velocity of the MOT is about 11~m/s. To load the most molecules, they should reach this capture velocity just as they arrive in the MOT capture volume.  If they reach low velocity prematurely, they diverge too much and are less likely to be captured. We turn on the slowing light at $t_{\rm on} = 2.5$~ms and apply a linear frequency chirp of $\alpha = 24.5$~MHz/ms between $t_{\rm chirp} = 3.4$~ms and $t=t_{\rm off}$, when the slowing light is turned off. The MOT does not load until the slowing light is turned off because this light pushes trappable molecules out of the trapping region. Figure \ref{fig:MOTLoading}(a) shows the fluorescence of molecules in the MOT region as a function of time, recorded by a PMT. For these data, $t_{\rm off} = 15$~ms, shown by the vertical dotted line. The orange curve shows the arrival time distribution when there is no slowing applied. The most probable arrival time is $8$~ms, corresponding to a speed of $150$~m/s. The blue curve shows the distribution when the slowing is applied but no molecules are captured because only one MOT beam is used and the magnetic field is off. At early times, up until 7.3~ms, the slowed and unslowed curves are similar because the fastest molecules ($>165$~m/s) do not interact with the slowing light. Then there is a dip in the slowed signal, followed by a broad bump at later times, corresponding to molecules that have been slowed. The dashed purple curve is from a simulation of the slowing that uses the experimental slowing parameters and free flight velocity distribution as inputs. No scaling is applied, and the result matches perfectly with the experiment, showing that we have a good understanding of the slowing. These simulations have also been validated previously~\cite{Truppe2017}. The red curve shows the distribution when the MOT magnetic field is on and all six beams are present. This curve roughly follows the blue curve until $t=t_{\rm off}$. Then there is a rise in signal as molecules are loaded into the MOT, followed by a decline as they are gradually lost. Figure \ref{fig:MOTLoading}(b) compares experiment and simulation results for MOT loading. The red curve is the same experimental data as shown in (a). The black curve shows the simulated arrival-time distribution of molecules that arrive at the MOT within a 1~cm diameter disk, with forward speeds below $10$~m/s, and arrival times greater than $t_{\rm off}$, which we call $N_{\rm trappable}(t)$. Without modelling any of the details of the capture process itself, we can estimate the number of trapped molecules to be $[\int_{t_{\rm off}}^{t} N_{\rm trappable}(t')\,dt'] e^{-(t-t_{\rm off})/\tau}$ where $\tau = 95$~ms is the measured lifetime (see section \ref{sec:lifetime}). This result is shown by the purple curve, whose height has been scaled to give a good match with the experiment. We see that this simulated MOT loading curve is in excellent agreement with the measured one. This analysis gives a clear picture of which molecules in the beam are captured by the MOT, which is especially important in designing strategies to increase the number of molecules loaded.

\begin{figure}
\centering
\includegraphics[scale=0.8]{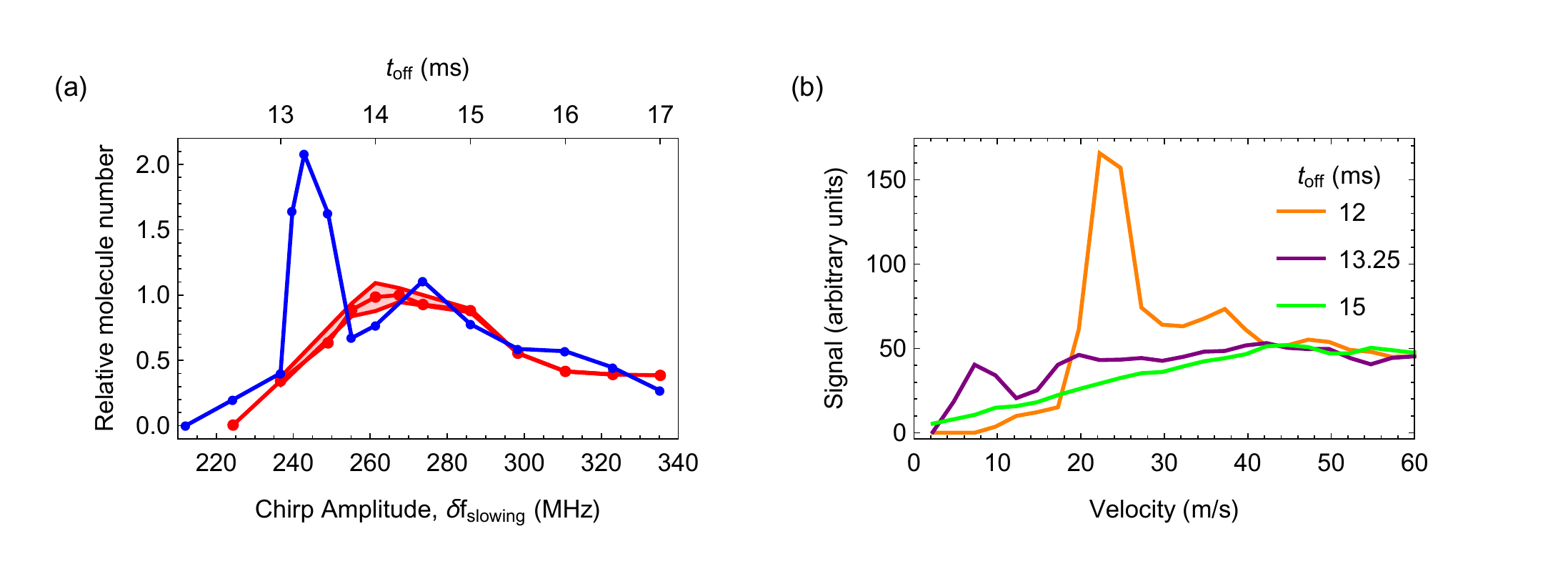}
\caption{(a) Relative number of trapped molecules versus $t_{\rm off}$ (or $\delta f_{\rm slowing}$, see text). Each point is obtained from 50 images averaged together. Red: experiment, with shaded band indicating typical fluctuations in the 50-image average; Blue: simulation. (b) Simulated velocity distributions for three different values of $t_{\rm off}$, 12~ms (orange), 13.25~ms (purple) and 15~ms (green).}
\label{fig:chirpPlots}
\end{figure}

Next we investigate how the number of molecules captured in the MOT depends on the chirp amplitude of the slowing laser, which is a key parameter of the slowing process. The red curve in figure \ref{fig:chirpPlots}(a) shows the measured effect of $t_{\rm off}$, which we also express in terms of the total frequency change $\delta f_{\rm slowing} = \alpha (t_{\rm off}-t_{\rm chirp})$. Increasing $t_{\rm off}$ reduces the final velocity of the slowest molecules, so the number of trapped molecules initially increases with $t_{\rm off}$. However, if $t_{\rm off}$ is too large many of the molecules reach zero velocity before they arrive at the MOT and so the trapped number falls again. The number in the MOT is largest when $t_{\rm off}=14.2$~ms, corresponding to $\delta f_{\rm slowing} = 265$~MHz. The coloured band indicates the fluctuations in the 50-image averages (the shot-to-shot fluctuations are 7 times larger). These fluctuations are large for $t_{\rm off}=14.2$~ms, so we prefer to use $t_{\rm off}=15$~ms [$\delta f_{\rm slowing}=285$~MHz] where the fluctuations are far smaller while the average number is only a little less. This is the value of  $t_{\rm off}$ we use for all subsequent data. The blue points in figure \ref{fig:chirpPlots}(a) show the expected number of molecules in the MOT for various $t_{\rm off}$ predicted by the slowing simulations discussed above in the context of figure \ref{fig:MOTLoading}. As before, we sum up all molecules that arrive at the MOT within a 1~cm diameter disk, with forward speeds below $10$~m/s, and arrival times greater than $t_{\rm off}$. These simulations show the same overall trends as in the experiment, but predict a large peak when $t_{\rm off} = 13.25$~ms which we do not see in the experiment. To understand this peak, it is helpful to study the velocity distributions found from the simulations. Figure \ref{fig:chirpPlots}(b) shows these velocity distributions for three values of $t_{\rm off}$. When $t_{\rm off} = 12$~ms, the velocity distribution has a narrow peak at 24~m/s, which are molecules that have faithfully followed the chirp, and then a broader, faster distribution which are those that have fallen behind the chirp. For this value of $t_{\rm off}$, there are hardly any molecules slow enough to be captured. Increasing $t_{\rm off}$ pushes the narrow peak in the velocity distribution to lower velocities. The peak also gets smaller because of the increased divergence of the slowest molecules.  When $t_{\rm off}=13.25$~ms, the narrow peak is pushed below 10~m/s and can be captured, producing the predicted peak in figure \ref{fig:chirpPlots}(a) for this $t_{\rm off}$. For $t_{\rm off} = 15$~ms, molecules in the slow peak are brought to rest or even turned around before reaching the MOT. However, some of the molecules which have fallen behind the chirp are now slow enough to be captured, so there is no sharp cut-off as $t_{\rm off}$ increases. At present, we do not know why we fail to observe the strong response predicted at $t_{\rm off}=13.25$~ms. The slowing simulations appear to be reliable~\cite{Truppe2017}, and they predict the observed arrival time distributions accurately [see figure \ref{fig:MOTLoading}]. We note that the larger fluctuations near the expected optimum $t_{\rm off}$ shows that a larger number is {\it sometimes} obtained, and suggests strong sensitivity to some parameter that is inadequately controlled.

\section{Properties of the MOT}

In this section we show how the properties of the MOT vary with the key parameters of the setup, mainly the total intensity at the MOT ($I_{00}$) of ${\cal L}_{00}$, the detuning ($\Delta_{00}$) of ${\cal L}_{00}$, and the axial magnetic field gradient ($dB/dz$). When not being varied these are set to be $I_{00}^0=400$~mW/cm$^2$, $\Delta_{00}^0=-0.75~\Gamma$ and $dB/dz=30.6$~G/cm. These standard parameters are also used for fluorescence imaging, unless otherwise stated.\\

\subsection{Number of molecules}

\begin{figure}
	\centering
	\includegraphics[scale=0.7]{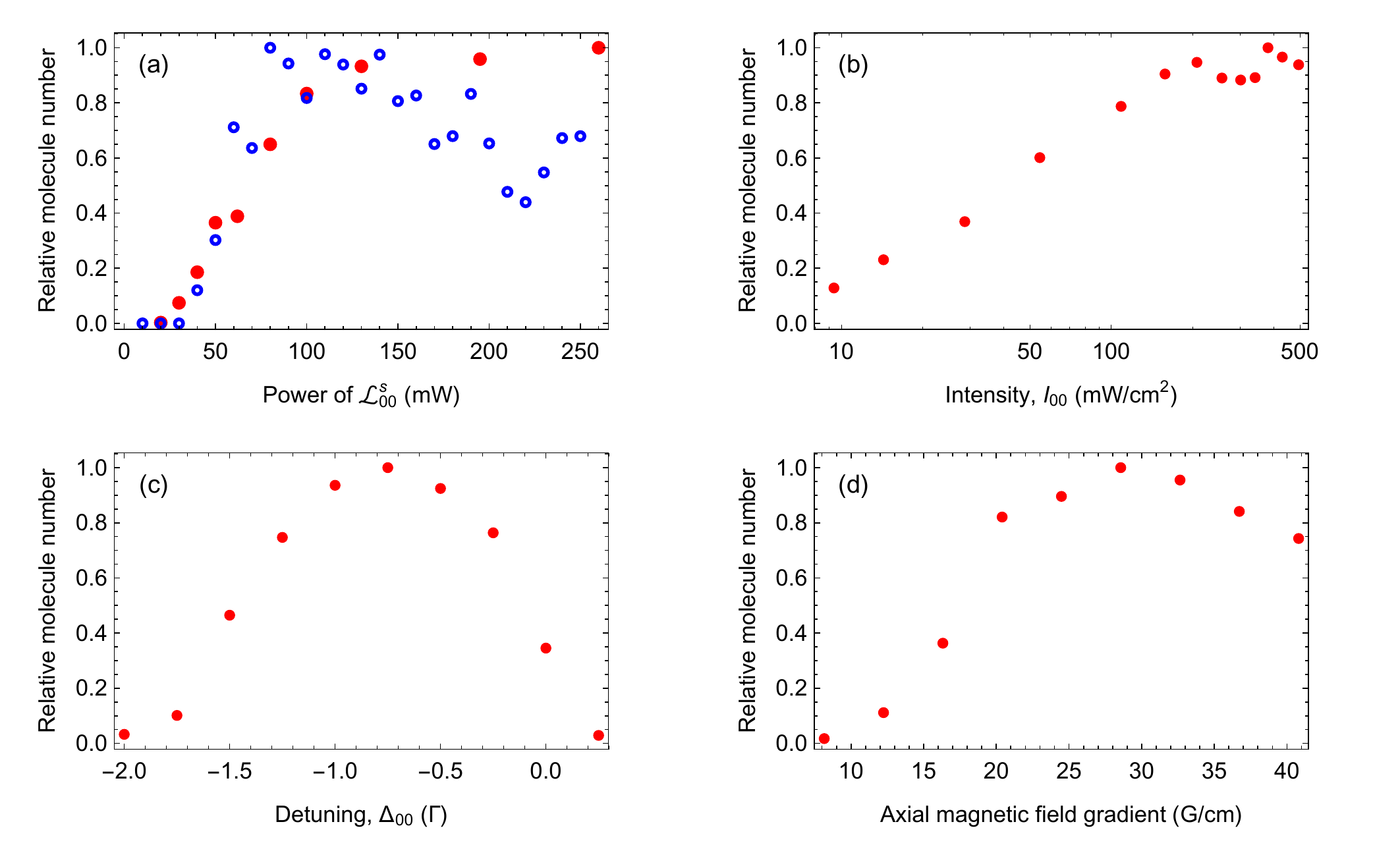}
	\caption{Relative number of molecules trapped as a function of (a) power of slowing laser ${\cal L}_{00}^{\rm s}$, (b) $I_{00}$ (c) $\Delta_{00}$ and (d) $dB/dz$. The uncertainty in the number due to shot-to-shot fluctuations is $3\%$. The MOT is imaged at $60$~ms with an exposure time of $30$~ms and averaged over 50 shots. In (a), the filled points are measurements while the open points are from slowing simulations.}
	\label{fig:numberPlots}
\end{figure}

We first investigate how to maximize the number of molecules in the MOT. The filled points in figure \ref{fig:numberPlots}(a) show how this number depends on the power of ${\cal L}_{00}^{\rm s}$. For powers below $20$~mW we observe no molecules in the MOT. Above this threshold the number increases roughly linearly until it saturates at a power near $120$~mW. The open points are the results of simulations identical to those discussed above but done for various ${\cal L}_{00}^{\rm s}$ powers. The simulations predict the same trends as we observe, with two notable exceptions. First, they predict that the trapped number will fall above a certain power, because at high power the molecules reach low speed too early. We do not see this effect in the experiment. Second, the simulations exhibit high sensitivity to the exact power (the sharp structures are not noise). The experimental data does not have the resolution to see this. We have also investigated how the number of molecules depends on the power of ${\cal L}_{10}^{\rm s}$. The nominal power is 130~mW, and we see no difference if we halve or double this value. 

Figure \ref{fig:numberPlots}(b) shows the relative number of molecules as a function of $I_{00}$. The number increases with intensity until 200~mW/cm$^2$, above which it remains constant within $10\%$. We note that we still load 10\% of this maximum number when $I_{00}$ is only 2\% of $I_{00}^0$. Figure \ref{fig:numberPlots}(c) shows the relative number of molecules versus $\Delta_{00}$. This shows a parabolic dependence with a maximum at $\Delta_{00}=-0.75\Gamma=-2\pi \times 6.2$~MHz. No MOT is formed when the light is blue-detuned or when it is red-detuned by more than $1.8\Gamma$. Figure \ref{fig:numberPlots}(d) shows the relative number of molecules versus $dB/dz$. We observe a MOT once $dB/dz > 10$~G/cm, and obtain the most molecules when $dB/dz=30$~G/cm. As the field gradient is increased beyond this, there is a slow decline in the number of trapped molecules, suggesting that the trap capture volume starts to decrease at these higher gradients.

\subsection{Scattering Rate}
\label{sec:scatteringRate}

A simple rate model~\cite{Tarbutt2013} can be used to predict the photon scattering rate of the molecules, and hence many of the properties of the MOT, as was done previously~\cite{Norrgard2016}. In this model, $n_g$ ground states are coupled to $n_e$ excited states, and the steady-state scattering rate is found to be
\begin{equation}
R_{\rm{sc}} = \Gamma \frac{n_e}{(n_g + n_e) + 2\sum_{j=1}^{n_{g}}(1+4\Delta_{j}^{2}/\Gamma^{2})I_{{\rm s},j}/I_{j}}.
\label{eq:Rsc1}
\end{equation}
Here, $I_j$ is the intensity of the light driving transition $j$, $\Delta_j$ is its detuning, and $I_{{\rm s},j}=\pi h c \Gamma/(3 \lambda_{j}^{3})$ is the saturation intensity for a two-level system with transition wavelength $\lambda_{j}$. For our MOT there are $n_g=24$ Zeeman sub-levels of the $v=0$ and $v=1$ ground states, all coupled to the same $n_{e}=4$ levels of the excited state. It is safe to neglect the $v=2$ and $v=3$ ground states since their populations are always small. In the experiment, the intensity of ${\cal L}_{10}$ is always higher than that of ${\cal L}_{00}$, and ${\cal L}_{10}$ is on resonance whereas ${\cal L}_{00}$ is detuned. It follows that the transitions driven by ${\cal L}_{00}$ dominate in the sum and we can neglect those driven by ${\cal L}_{10}$. The transitions driven by ${\cal L}_{00}$ have common values for $\Delta$ and $I_{\rm{s}}$, and the total intensity, $I_{00}$, is divided roughly equally between them so that we can write $I_{j}=I_{00}/12$. With these approximations, the scattering rate becomes

\begin{equation}
\centering
R_{\rm sc}=\frac{\Gamma_{\rm {eff}}}{2} \frac{\rm{s_{eff}}}{1+s_{\rm{eff}}+4\Delta^2/{\Gamma^2}},
\label{eq:Rsc}
\end{equation}
where 
\begin{equation}
\centering
\Gamma_{\rm eff}=\frac{2n_{\rm e}}{n_{\rm g}+n_{\rm e}}\Gamma=\frac{2}{7}\Gamma,
\label{eq:Gamma}
\end{equation}
and
\begin{equation}
\centering
s_{\rm eff}=\frac{I_{00}}{I_{\rm s,eff}}=\frac{2(n_{\rm g}+n_{\rm e})}{n^2_{\rm g}}\frac{I_{00}}{I_{\rm s}}.
\label{eq:seff}
\end{equation}
Equation (\ref{eq:Gamma}) gives $\Gamma_{\rm eff} = 14.9 \times 10^{6}$~s$^{-1}$, and equation (\ref{eq:seff}) gives $I_{\rm s,eff} = 50~{\rm mW/cm}^{2}$.

\begin{figure}
	\centering
	\includegraphics[scale=0.7]{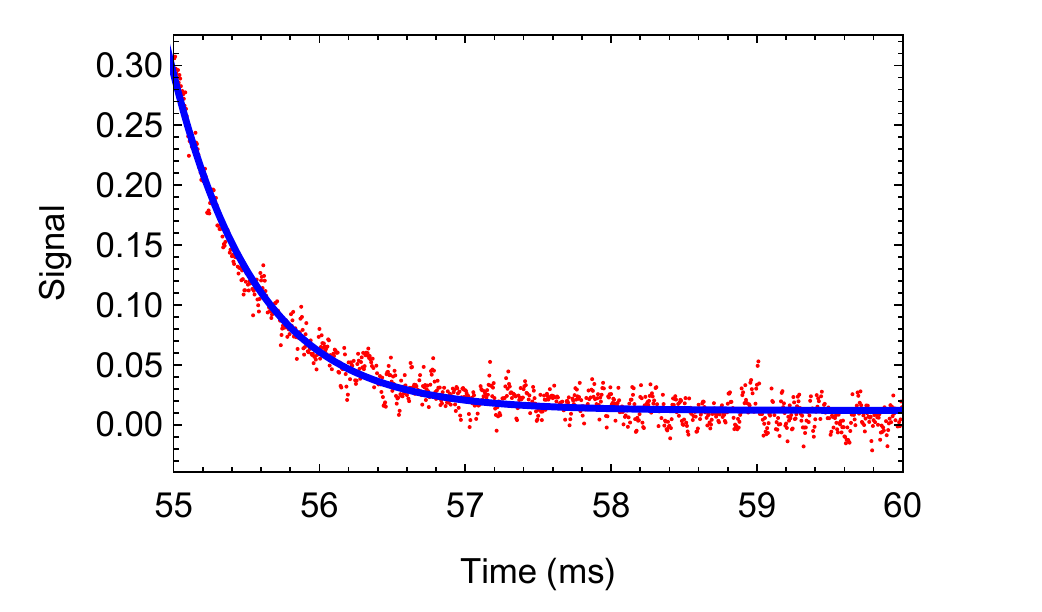}
	\caption{Dots: MOT fluorescence as a function of time after turning ${\cal L}_{21}$ off at $t=55$~ms. Line: fit to a single exponential decay, giving a time constant $\tau_{2}=572(7)~\mu$s.}
	\label{fig:scatterRateMeasurement}
\end{figure}

We measure the scattering rate in the MOT by turning off ${\cal L}_{21}$ and detecting the decay of fluorescence as molecules are optically pumped into $v=2$. The scattering rate is simply $R_{\rm sc} = 1/(b_{2}\tau_{2})$, where $b_2$ is the branching ratio for the excited state to decay to $v=2$, and $\tau_{2}$ is the measured $1/e$ decay constant of the fluorescence. We determine $b_{2}$ by comparing the MOT fluorescence intensity on the $A(0) \rightarrow X(0)$ ($\lambda_{00} =606$~nm) and $A(0) \rightarrow X(2)$ ($\lambda_{20} = 652$~nm) transitions, where we are using the same notation as in Table~\ref{tab:lasers}. We isolate these two contributions to the fluorescence using bandpass filters placed between the two lenses of the imaging system, where the light from the MOT is collimated. We switch back and forth frequently between the two filters. Each time the filter is switched we take an image with no molecules present, and then subtract this from the MOT image so that only the fluorescence from the molecules remains. The pass band of each filter has a full width at half maximum of 20~nm, and the transmission exceeds 93\%. Importantly, the filter that transmits at $\lambda_{20}$ has a transmission of less than $10^{-6}$ at $\lambda_{00}$, which is small enough to neglect. We can express $b_{2}$ in terms of known quantities as follows:
\begin{equation}
b_{2} = b_{0} \frac{I_2}{I_0}\frac{T_{{\rm lens 1}}^{\lambda_{00}}}{T_{{\rm lens 1}}^{\lambda_{20}}}\frac{T_{{\rm lens 2}}^{\lambda_{00}}}{T_{{\rm lens 2}}^{\lambda_{20}}}\frac{T_{{\rm window}}^{\lambda_{00}}}{T_{{\rm window}}^{\lambda_{20}}}\frac{T_{{\rm filter 0}}^{\lambda_{00}}}{T_{{\rm filter 2}}^{\lambda_{20}}} \frac{\epsilon_{{\rm camera}}^{\lambda_{00}}}{\epsilon_{{\rm camera}}^{\lambda_{20}}}.
\end{equation} 
Here, $I_{2}/I_{0}$ is the measured fluorescence ratio, $b_{0}=0.987\pm ^{0.013}_{0.019}$ is the branching ratio to $v=0$~\cite{Wall2008}, $T_{{\rm lens 1(2)}}^{\lambda}$ is the transmission of lens 1(2), $T_{{\rm window}}^{\lambda}$ is the transmission of the vacuum viewport, $T_{{\rm filter 0(2)}}^{\lambda}$ is the transmission  of the filter that isolates the fluorescence to $v=0(2)$, and $\epsilon_{{\rm camera}}^{\lambda}$ is the quantum efficiency of the camera, all at wavelength $\lambda$. All transmissions and quantum efficiencies are taken from data supplied by the manufacturers. The result is $b_{2} = 8.4(5)\times 10^{-4}$. This is 40\% smaller than the theoretical value of $1.2\times 10^{-3}$ given in reference \cite{Pelegrini2005}. Figure \ref{fig:scatterRateMeasurement} shows an example of a scattering rate measurement where the time constant is $\tau_{2}=572(7)~\mu$s giving a scattering rate of $2.08(13) \times 10^6$~s$^{-1}$

\begin{figure}
\centering
\includegraphics[scale=0.7]{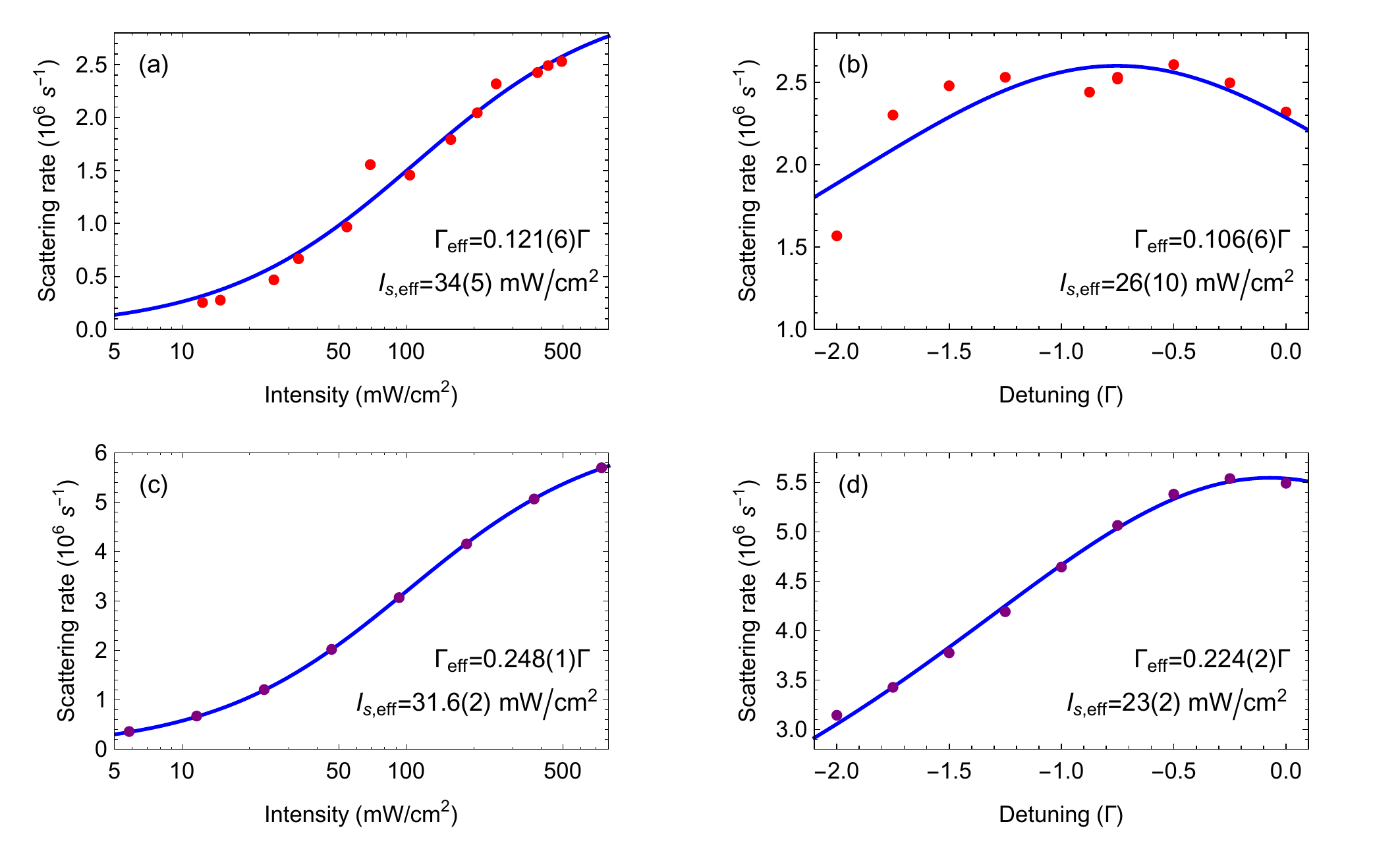}
\caption{Scattering rate versus intensity and detuning of ${\cal L}_{00}$: (a,b) measured, (c,d) modelled. Solid lines are fits to equation (\ref{eq:Rsc}), and the fit parameters are given in each case.}
\label{fig:scatterPlots}
\end{figure}

Figures \ref{fig:scatterPlots}(a) and (c) show the measured and simulated scattering rate versus $I_{00}$, along with fits to equation (\ref{eq:Rsc}) and the associated fit parameters. Both the experimental and simulated results fit well to this model. The values of $I_{\rm s,eff}$ are in agreement with each other and are close to that predicted by the simple model outlined above. The simulation gives $\Gamma_{\rm eff}$ close to the predicted value of $0.29\Gamma$, but the experimental value is a factor of 2 smaller than simulated. This difference could be caused by optical pumping into coherent states that, for short periods, are dark to the MOT light, an effect which cannot be captured by the rate model. There are no stable dark states for molecules that move quickly enough through the non-uniform magnetic field and light polarization, but there may be states that are decoupled from the light for long enough to limit the scattering rate. Figure \ref{fig:scatterPlots}(b) and (d) show the measured and simulated scattering rate versus $\Delta_{00}$, along with fits to equation (\ref{eq:Rsc}). The fit to the measurements is unconvincing, though the fit parameters are consistent with those found in (a), showing that equation (\ref{eq:Rsc}) is sufficient to represent the dependence of the scattering rate on both intensity and detuning. The simulated data fits well to the model but with parameters somewhat different to those found in (c). 

\subsection{Oscillation frequency and damping constant}

We consider the motion of the molecules in the direction of the slowing laser, since it is straightforward to give the molecules a push in this direction, and because the camera views this axis. Consider a molecule with displacement $x$ and velocity $v$ along this axis, interacting with the six MOT beams. At low intensity, $s_{\rm eff} \ll 1$, the total scattering rate is the sum of the scattering rates from each beam individually, so it is easy to identify the force exerted by a single beam. This is not the case at higher intensities. Nevertheless, the force due to one of the MOT beams in the horizontal plane may be written as
\begin{equation}
F_{\pm}(x,v) = \mp \frac{\hbar k}{\sqrt{2}} \frac{\Gamma_{\rm eff}}{2}\frac{s_{\rm eff}/6}{1+s_{\rm eff}+4(\Delta \pm \frac{k v}{\sqrt{2}} \pm g_{\rm eff}\mu_{\rm B} A x/\hbar)^{2}/\Gamma^{2}}
\label{eq:force1DMOT}
\end{equation}
where $k=2\pi/\lambda$ is the wavevector, $\Delta$ is the laser detuning, $g_{\rm eff}$ is an effective magnetic g-factor for the transition, $A$ is the magnetic field gradient in the horizontal plane, $s_{\rm eff}$ is the saturation parameter for all six beams, and the factors of $\sqrt{2}$ account for the 45$^{\circ}$ angle between the MOT beams and the $x$-axis. Here, in the numerator we have used the intensity of the single beam applying the force, while in the denominator we have used the intensity of all six beams to account for the saturation of the scattering rate by the full intensity. This approximation is known to work well for the modest saturation parameters used in this work~\cite{Lett89}. The total force in the $x$-direction is 
\begin{equation}
F=2(F_{-} + F_{+}) = m x'',
\label{eq:totalForce}
\end{equation} 
where $m$ is the mass and the factor of 2 accounts for there being two pairs of horizontal MOT beams. Using a Taylor expansion about $x=x'=0$, we obtain
\begin{equation}
x'' = -\omega^{2}x - \beta x',
\label{eq:ho}
\end{equation} 
where $\omega$, the trap oscillation angular frequency, is given by
\begin{equation}
\omega^{2} = -\frac{4\sqrt{2}}{3m}\frac{\Gamma_{\rm {eff}}}{\Gamma} \frac{k (\Delta/\Gamma)g_{\rm eff}\mu_{\rm B} A s_{\rm eff}}{(1+s_{\rm eff}+4\Delta^{2}/\Gamma^{2})^{2}},
\label{eq:frequency}
\end{equation}
and $\beta$, the damping constant, is
\begin{equation}
\beta = -\frac{4}{3m}\frac{\Gamma_{\rm {eff}}}{\Gamma} \frac{\hbar k^{2} (\Delta/\Gamma) s_{\rm eff}}{(1+s_{\rm eff}+4\Delta^{2}/\Gamma^{2})^{2}}.
\label{eq:damping}
\end{equation}

\begin{figure}[t]
	\centering
	\includegraphics[scale=0.7]{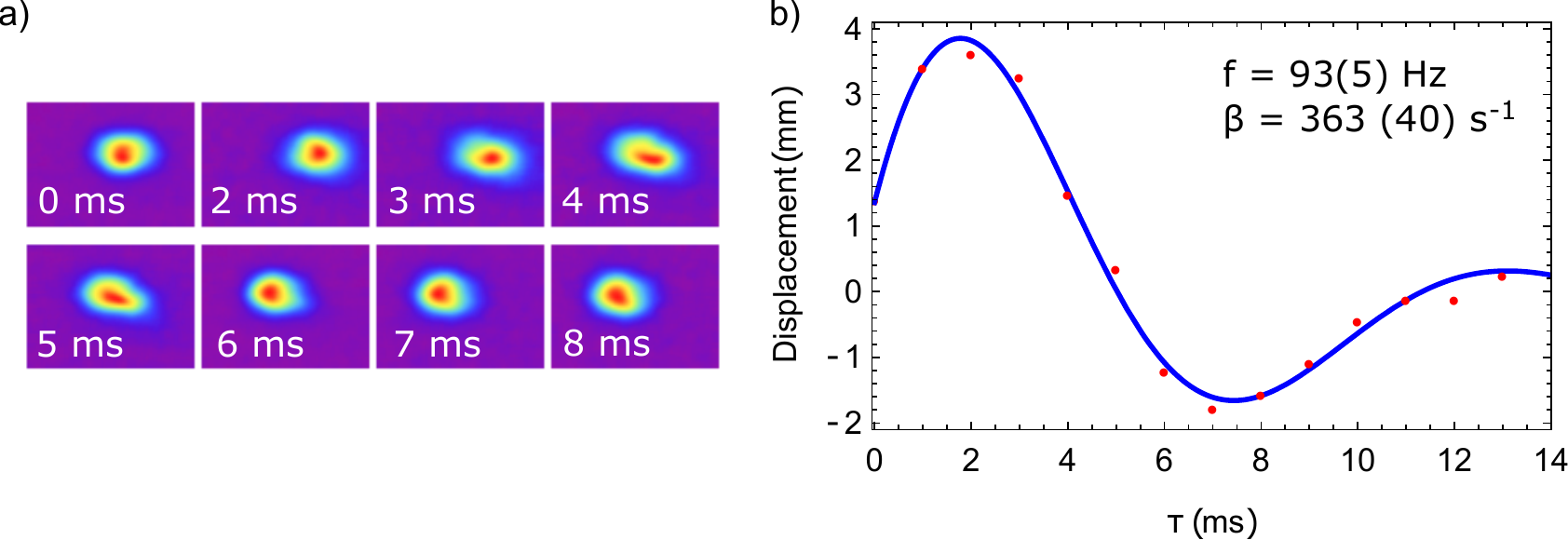}
	\caption{(a) Fluorescence images of the MOT at various times after a radial push. Each image is a 0.5~ms exposure averaged over 30 shots. (b) Radial displacement versus time after the push, showing the damped harmonic motion of the MOT. Here, $\Delta_{00}=-0.75\Gamma$ and $I_{00}=200$~mW/cm$^2$. Points: experimental data. Line: fit to equation (\ref{eq:dampedHarmonicOscillator}).}
	\label{fig:oscMethod}
\end{figure}

To measure the radial trap oscillation frequency, $f=\omega/(2\pi)$, and the damping constant, $\beta$, we pulse on ${\cal L}_{00}^{s}$ for 0.5~ms to push the cloud in the radial direction. We then image the cloud after various delay times using an exposure time of $0.5$~ms, which is short compared to the oscillation period. Figure \ref{fig:oscMethod}(a) is a sequence of such images showing the damped oscillation of the cloud. We integrate each image over the axial coordinate to give radial distributions, and then fit a Gaussian model to each distribution to obtain the central position of the cloud at each time. Figure \ref{fig:oscMethod}(b) shows the mean radial displacement versus time. A suitable solution of equation (\ref{eq:ho}) for the displacement from the equilibrium position $x_{0}$ is
\begin{equation}
x-x_0 = a e^{-\beta t/2}\cos(\sqrt{(\omega^2-\beta^2/4)}t-\phi),
\label{eq:dampedHarmonicOscillator}
\end{equation}
where $a$ is the amplitude of oscillation and $\phi$ is a phase. All the oscillation data presented here fit well to this model. Figure \ref{fig:oscMethod}(b) shows an example of this fit for the case where $\Delta_{00}=-0.75\Gamma$ and $I_{00}=200$~mW/cm$^{2}$. For these parameters, we find $f=93(5)$~Hz and $\beta=363(40)$~s$^{-1}$. Here, the uncertainties are the standard deviations of repeated measurements. 

\begin{figure}[t]
	\centering
	\includegraphics[scale=0.7]{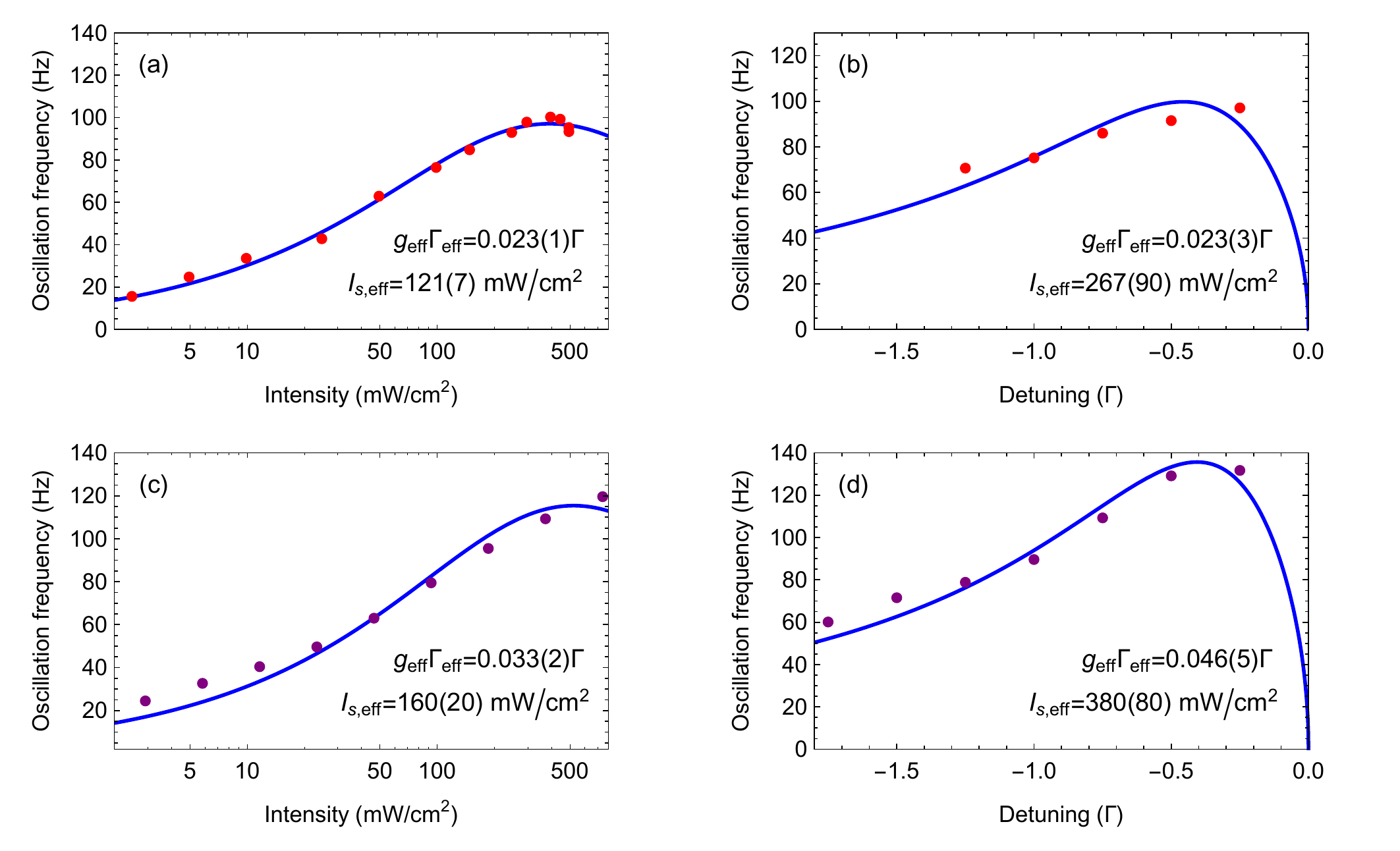}
	\caption{Oscillation frequency versus intensity and detuning of ${\cal L}_{00}$: (a,b) measured, (c,d) modelled. Solid lines are fits to $\omega/(2\pi)$ given by equation (\ref{eq:frequency}), and the fit parameters are given in each case. Typical fluctuations in repeat measurements are $5\%$.}
	\label{fig:oscPlots}
\end{figure}

Figures \ref{fig:oscPlots}(a,c) show the measured and simulated oscillation frequency as a function of $I_{00}$. The measured frequency increases with $I_{00}$ until it reaches a maximum at $400$~mW/cm$^2$. The measured and simulated frequencies are in excellent agreement, differing by less than 20\% across the whole range of intensities. We fit these data to the frequency given by equation (\ref{eq:frequency}). In this equation, all the parameters are known apart from $g_{\rm eff}$, $\Gamma_{\rm {eff}}$ and $I_{\rm s,eff}$. The first two appear only as a product, so the free parameters of the fit are $g_{\rm eff}\Gamma_{\rm {eff}}$, which is a measure of the strength of confinement, and $I_{\rm s,eff}$ which sets the intensity required to reach the strongest confinement. The measurements and simulations fit well to this simple model, with the best fit parameters given in the figure. They show that the saturation intensity appropriate to the oscillation frequency is considerably higher than that found for the scattering rate -- it takes more power to trap the molecules than might be expected from the scattering rate data. It is peculiar that the experimental and simulated oscillation frequencies are in such good agreement even though the scattering rates differ by a factor of 2. Figures \ref{fig:oscPlots}(b,d) show the measured and simulated oscillation frequency as a function of $\Delta_{00}$, along with fits to equation (\ref{eq:frequency}).  Both experiments and simulations suggest that the maximum $f$ occurs very close to resonance, requiring a large $I_{\rm s,eff}$ in equation (\ref{eq:frequency}) for a good fit. As with the scattering rate data, the best fit parameters obtained from the intensity dependence and detuning dependence are consistent with one another. It is interesting that, over the range of $\Delta$ explored here, the oscillation frequency shows a weak and simple dependence on $\Delta$ despite the small interval ($3\Gamma$) between the upper two hyperfine states.

\begin{figure}[t]
	\centering
	\includegraphics[scale=0.7]{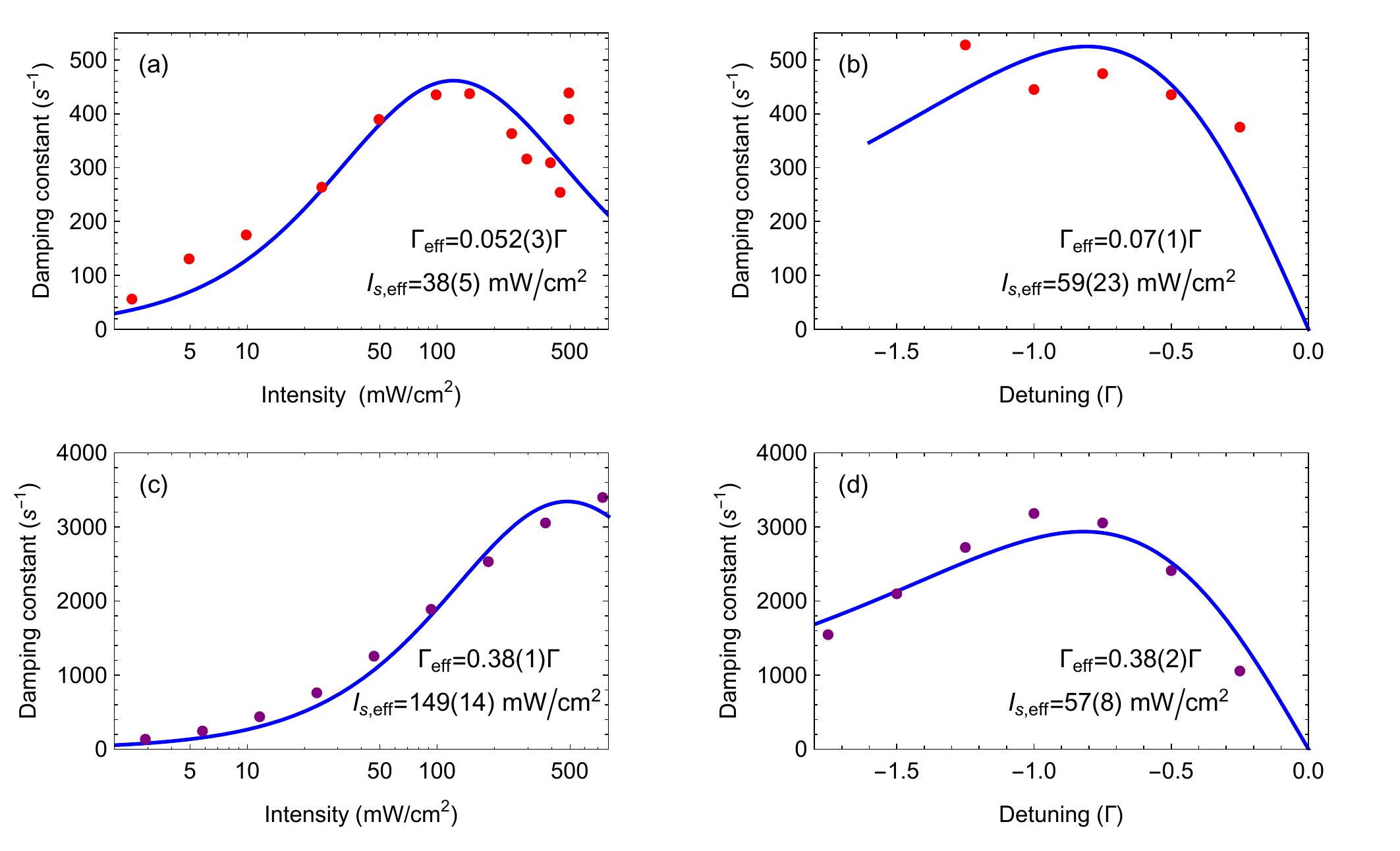}
	\caption{Damping constant versus intensity and detuning of ${\cal L}_{00}$: (a,b) measured, (c,d) modelled. Solid lines are fits to equation (\ref{eq:damping}), and the fit parameters are given in each case. Typical fluctuations in repeat measurements are $10\%$.}
	\label{fig:dampPlots}
\end{figure}

Figures \ref{fig:dampPlots}(a,c) show the measured and simulated damping constant as a function of $I_{00}$. We see that the measured values of $\beta$ are far smaller than the simulated ones across the whole intensity range. At low intensities, $I_{00} < 25$~mW/cm$^{2}$, the measured $\beta$ is 2--3 times smaller than simulated, while at high intensities the discrepancy is a factor of 5--10. In the experiment, $\beta$ has a maximum near 100~mW/cm$^{2}$, while in the simulations the maximum is beyond 750~mW/cm$^{2}$. Damping constants much smaller than the simulated ones are also found in both the dc and rf MOTs of SrF~\cite{McCarron2015, Norrgard2016}. We tentatively attribute this to the effect of the polarization gradient force. This force has the opposite sign to the Doppler cooling force, and dominates at low velocities, especially when the intensity is high~\cite{Devlin2016}. This could result in a reduced damping constant, with a higher reduction factor at higher intensities. We are currently investigating the role of these polarization gradient forces in the molecular MOT. Despite the discrepancy between experiment and simulation, both datasets follow equation (\ref{eq:damping}), as can be seen by the fits in figures \ref{fig:oscPlots}(a,c). The much weaker damping in the experiment is reflected by a much smaller $\Gamma_{\rm {eff}}$ in the fit, and the shift of the maximum to lower intensity in the experiment is reflected by a smaller $I_{\rm s,eff}$. Figures \ref{fig:dampPlots}(b,d) show the measured and simulated damping constant as a function of $\Delta_{00}$. We see that $\beta$ gradually decreases as $|\Delta_{00}|$ approaches zero, which is the opposite behaviour to $f$. Thus, the choice of detuning is a trade-off between maximizing $f$ and $\beta$. Unlike the dependence on $I_{00}$, the experimental dependence on $\Delta_{00}$ does not seem to follow equation (\ref{eq:damping}). The simulation results do roughly follow this equation however.

\subsection{Temperature}

The expected temperature can be expressed in terms of the damping constant $\beta$ and a velocity-independent momentum diffusion coefficient $D$, following a standard treatment \cite{Gordon1980} extended to three dimensions. For a force that is linear in the momentum, $\vec{F}=-\beta \vec{p}$, the cooling power is $P_{\rm cool}= \beta p^{2}/m = 2\beta E$, where $E$ is the kinetic energy. The heating power is $P_{\rm heat} = \frac{d}{dt}\frac{\langle p^{2}\rangle}{2m} = D/m$. Here, we have used the definition $2D = \frac{d}{dt}(\langle p^{2} \rangle - \langle p \rangle^{2})$, and the fact that $\langle p \rangle = 0$. Equating the heating and cooling powers we find an equilibrium energy $E=\frac{3}{2} k_{\rm B}T=\frac{D}{2m \beta}$. If the light field has no intensity gradients, so that there is no heating due to fluctuations of the dipole force, the momentum diffusion is due only to the two randomly-directed recoils per absorption-spontaneous emission cycle: 
\begin{equation}
2D = 2(\hbar k)^{2} R_{\rm sc}.
\label{eq:diffusionConstant}
\end{equation}
This gives us the temperature
\begin{equation}
T = \frac{1}{3} \frac{(\hbar k)^{2} R_{\rm sc}}{k_{\rm B}m\beta}.
\label{eq:expectedTemperature}
\end{equation}
Using equations (\ref{eq:Rsc}) and (\ref{eq:damping}) for $R_{\rm sc}$ and $\beta$ we obtain an expression for the Doppler temperature: 
\begin{equation}
T_{\rm D} = -\frac{\hbar\Gamma^{2}}{8 k_{\rm B} \Delta}(1+ s_{\rm eff} + 4\Delta^{2}/\Gamma^{2}).
\label{eq:DopplerTemperature}
\end{equation}
This is identical to the Doppler temperature for a two-level atom. We note that the expression for $D$ is modified slightly at intermediate intensities, and that fluctuations of the dipole force can alter $D$ considerably when there are intensity gradients~\cite{Gordon1980}. Nevertheless, equation (\ref{eq:DopplerTemperature}) has been verified for a three dimensional MOT in conditions where sub-Doppler processes are ineffective~\cite{Chang2014}.

\begin{figure}[tb]
	\centering
	\includegraphics[scale=0.6]{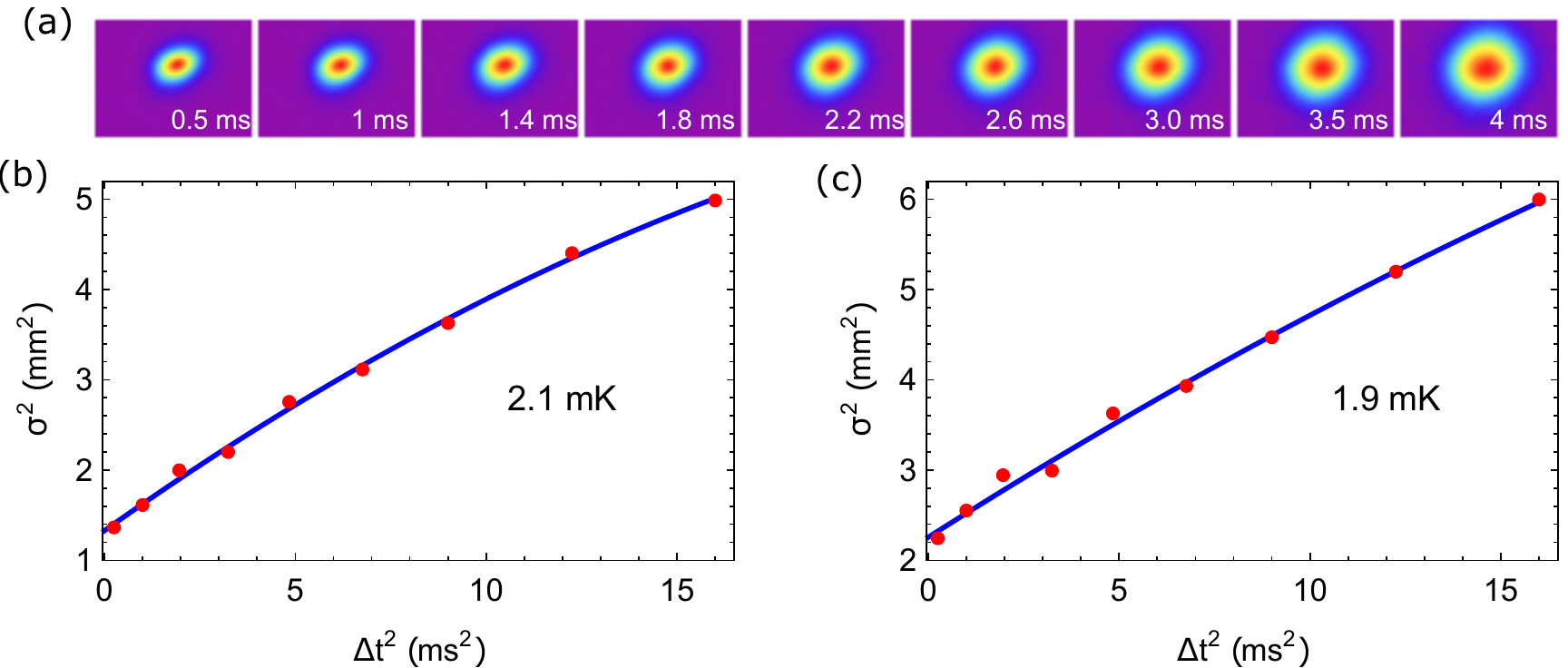}
	\caption{(a) Fluorescence images of expanding cloud after release from a MOT with  $I_{00}=40$~mW/cm$^2$. Each image is a 1~ms exposure and is averaged over 20 shots. (b,c) Points show $\sigma^2$ versus $\Delta t^2$ in the axial and radial directions respectively. Lines are fits to the model discussed in the text and give temperatures of 2.1~mK and 1.9~mK respectively.}
	\label{fig:tempMethod}
\end{figure}

To measure the temperature of the MOT we record fluorescence images after various free expansion times. We first load the MOT with the maximum number of molecules by using the intensity $I^0_{00}$ and detuning $\Delta^0_{00}$. At $t=50$~ms we either ramp the intensity to a new value over a period of $20$~ms or we jump the detuning to a new value. At $t=75$~ms we turn off the magnetic field and ${\cal L}_{00}$ so that the cloud is free to expand. After a free expansion time $\Delta t$, ${\cal L}_{00}$ is turned back on with the standard parameters $I^0_{00}$ and $\Delta^0_{00}$, and the fluorescence is imaged for 1~ms. Figure \ref{fig:tempMethod}(a) shows the expansion of the cloud in a sequence of images for several $\Delta t$. Each image is the sum of 20 repeats of the measurement. For each image, we sum over the axial (radial) coordinate to obtain the radial (axial) distribution. To each distribution we fit the Gaussian model $n(x) = A e^{-(x-x_0)^{2}/(2\sigma^{2})}$. Figures \ref{fig:tempMethod}(b,c) show $\sigma^{2}$ versus $(\Delta t)^{2}$ for the axial and radial directions. For free expansion, the rms width $\sigma$ follows $\sigma^2=\sigma^2_0+k_B T(\Delta t)^2/m$, where $\sigma_0$ is the initial rms width and $T$ is the temperature. Molecules in the wings of the cloud scatter at a slightly lower rate than those at the centre due to the change of laser intensity across the cloud. This slightly reduces the apparent size of the cloud, and since the effect is stronger for larger clouds the relation between $\sigma^{2}$ and $(\Delta t)^{2}$ becomes slightly non-linear. We account for this by fitting the data to $\sigma^2=\sigma^2_0+k_B T(\Delta t)^2/m + a (\Delta t)^{4}$. A model of fluorescence imaging that takes this effect into account verifies that this approach gives reliable temperatures~\cite{Truppe2017b}. For our data, inclusion of the $(\Delta t)^{4}$ term in the fit typically gives a temperature about 10\% higher than otherwise. Other potential systematic errors in these temperature measurements were considered in \cite{Truppe2017b} and found to be negligible. Figures \ref{fig:tempMethod}(b,c) show examples of the fit where the axial and radial temperatures are found to be $T_{z}=2.1$~mK and $T_{\rho}=1.9$~mK. The temperatures for the two directions are always close, so we take the geometric mean $T=T_{\rho}^{2/3}T_{z}^{1/3}$.

\begin{figure}[tb]
\centering
\includegraphics[scale=0.7]{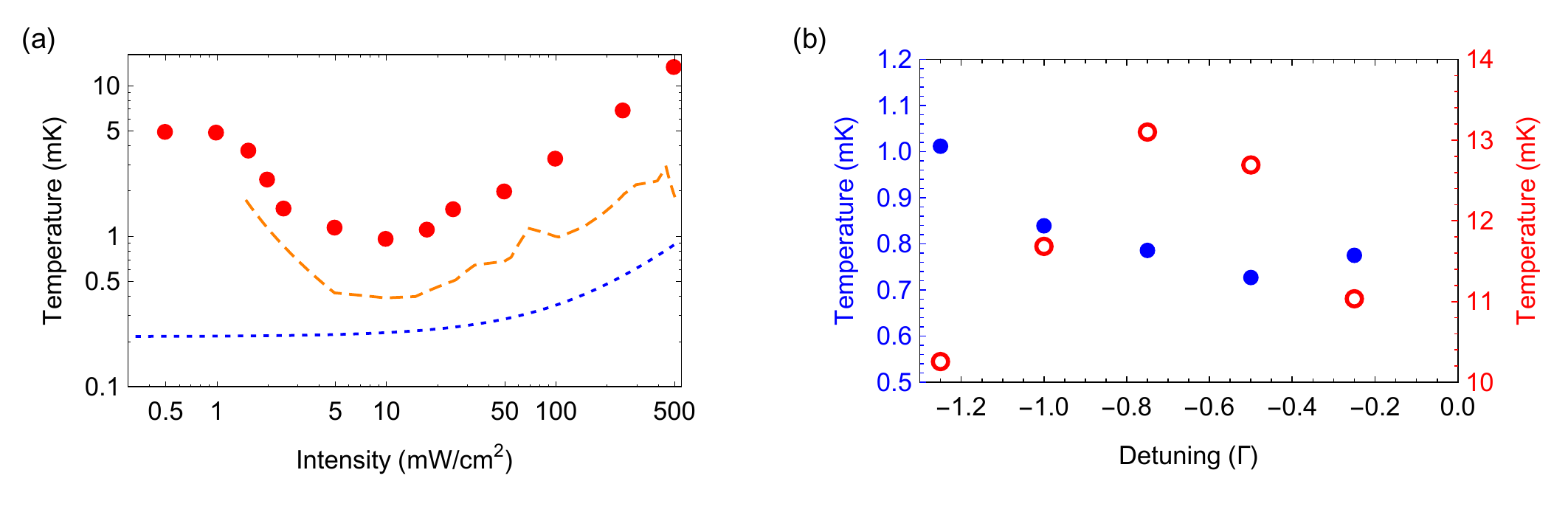}
\caption{(a) Temperature versus $I_{00}$. Points: measurements. Dotted blue line: Doppler temperature given by equation (\ref{eq:DopplerTemperature}). Dashed orange line: Temperature calculated using equation (\ref{eq:expectedTemperature}) along with the measured scattering rate and damping constant. (b) Temperature versus $\Delta_{00}$ for two different values of $I_{00}$, $400$~mW/cm$^2$ (red, open points, right hand scale) and $4$~mW/cm$^2$ (blue, filled points, left hand scale). Each data point is the temperature determined from a sequence of nine images using the method illustrated in figure \ref{fig:tempMethod}.}
\label{fig:tempPlots}
\end{figure}

Figure \ref{fig:tempPlots}(a) shows the temperature versus $I_{00}$ together with the Doppler temperature given by equation (\ref{eq:DopplerTemperature}). At full intensity ($I_{00}^0$), the temperature is 13~mK, which is 17 times higher than expected from equation (\ref{eq:DopplerTemperature}). The temperature decreases as the intensity decreases, reaching a minimum at $9$~mW/cm$^2$ where it is $960$~$\mu$K, 4 times the Doppler temperature. When the intensity is reduced below $9$~mW/cm$^2$, the temperature increases again. According to equation (\ref{eq:expectedTemperature}) the temperature is related in a simple way to the scattering rate and the damping constant. Since we have measured $T$, $R_{\rm sc}$ and $\beta$ across a wide range of intensities, we can test whether this relation is accurate for our MOT. Using linear interpolations over the measured values of $R_{\rm sc}$ [figure \ref{fig:scatterPlots}(a)] and $\beta$ [figure \ref{fig:dampPlots}(a)] at various intensities, and equation (\ref{eq:expectedTemperature}), we obtain the expected temperature shown by the dashed line in figure \ref{fig:tempPlots}(a). This shows exactly the same intensity dependence as we measure. The temperature found from equation (\ref{eq:expectedTemperature}) is higher than the Doppler temperature because the damping constant is smaller than predicted. The measured temperature is higher again, by a factor of 2 for intensities below 10~mW/cm$^{2}$ and by a factor of 3 at higher intensities. This shows that the diffusion constant is higher than that given by equation (\ref{eq:diffusionConstant}). This might be due to dipole force fluctuations which are not included in equation (\ref{eq:diffusionConstant}). We note that excess heating at high intensity is also seen in atomic MOTs \cite{Xu2003, Kemp2016} and various explanations have been given such as the effect of coherences between excited state sub-levels~\cite{Choi2008} and transverse intensity fluctuations of the MOT beams~\cite{Chaneliere2005}. 

Figure \ref{fig:tempPlots}(b) shows how the temperature depends on $\Delta_{00}$ at both high intensity (400~mW/cm$^2$) and low intensity (4~mW/cm$^2$). At high intensity the temperature is highest at $\Delta_{00}=-0.75$~$\Gamma$ and decreases at both larger and smaller detunings. At low intensity the temperature decreases as $|\Delta_{00}|$ decreases, until $\Delta_{00}=-0.5$~$\Gamma$ where it reaches a minimum of 730~$\mu$K, 3.5 times the Doppler temperature. Our previous work shows how to reduce the temperature below the Doppler limit using a blue-detuned optical molasses~\cite{Truppe2017b}.

\subsection{Cloud size}

\begin{figure}
	\centering
	\includegraphics[scale=0.7]{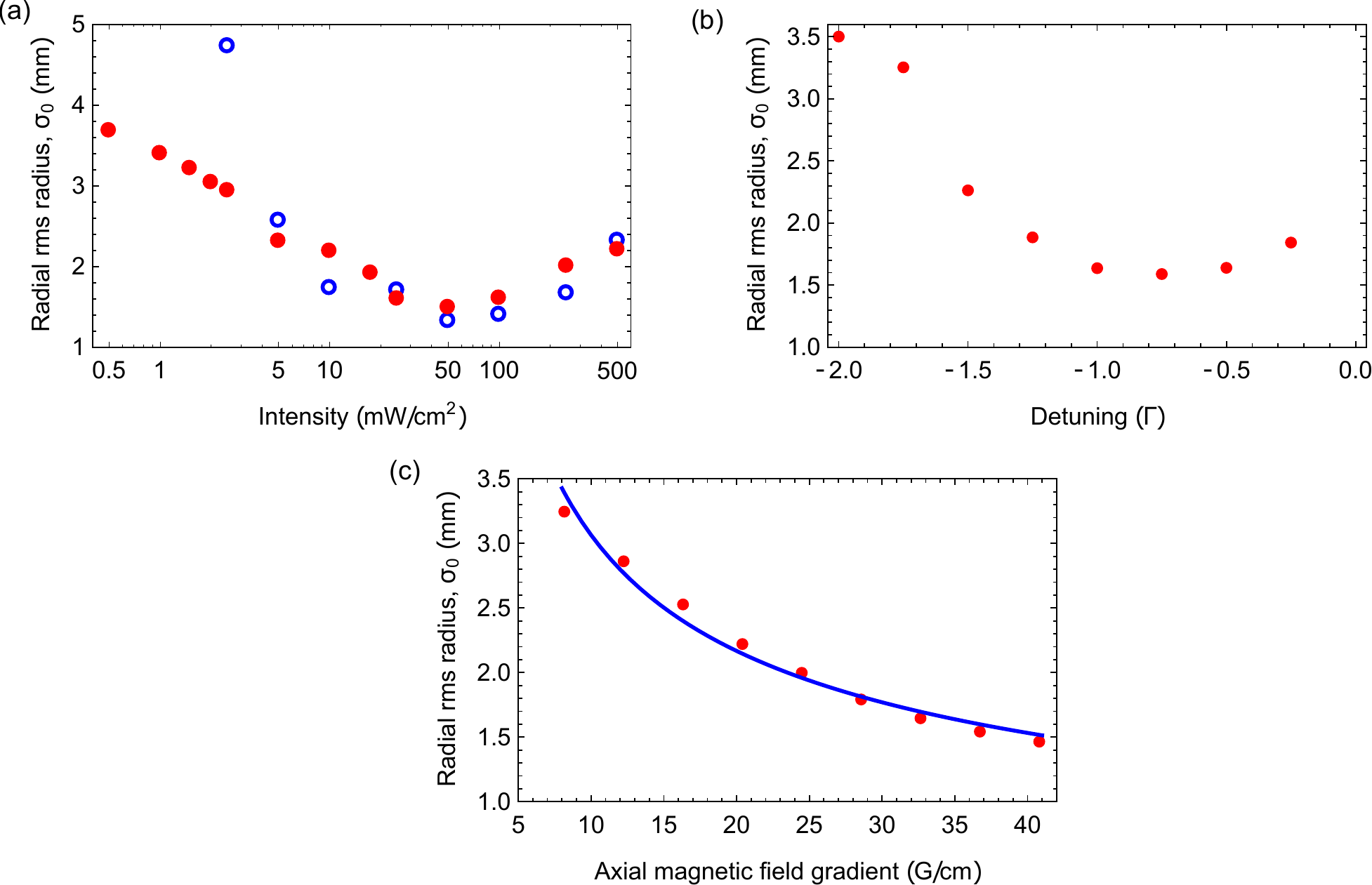}
	\caption{(a) Radial size of MOT versus intensity, $I_{00}$, showing measurements (filled, red points) and the prediction of equation (\ref{eq:equipartition}) (open blue points). (b) Radial size of MOT versus detuning, $\Delta_{00}$. (c) Radial size of MOT versus axial magnetic field gradient. Points: measurements. Line: fit to $\sigma_{0} = \zeta (dB/dz)^{-1/2}$. In (a) the values of $\sigma_0$ are obtained from the same ballistic expansion data that give the temperatures in figure \ref{fig:tempPlots}(a). In (b) and (c) the size is obtained by imaging the MOT directly for various values of $\Delta_{00}$ and $dB/dz$.}
	\label{fig:sizePlots}
\end{figure}

The filled points in figure \ref{fig:sizePlots}(a) show the rms radial size of the cloud, $\sigma_{0}$, as a function of $I_{00}$. We can interpret these data with the help of the equipartition theorem which relates $\sigma_0$ to $\omega$ and $T$:
\begin{equation}
\sigma_{0} = \sqrt{\frac{k_B T}{m \omega^2}}.
\label{eq:equipartition}
\end{equation} 
As $I_{00}$ is reduced from 500 to 50~mW/cm$^{2}$ the cloud size decreases. This is because $T$ falls by a factor of 7 over this intensity range [see figure \ref{fig:tempPlots}], whereas $\omega$ only falls by 50\% [see figure \ref{fig:oscPlots}]. As $I_{00}$ decreases further the cloud size increases because $\omega$ falls while $T$ stops falling and then starts increasing. The open points in figure \ref{fig:sizePlots}(a) are the predictions of equation (\ref{eq:equipartition}) for those values of $I_{00}$ where we have measured both $\omega$ and $T$. These predictions agree well with the measurements between 5 and 500~mW/cm$^{2}$. At lower intensity, the measured size is smaller than predicted by equation (\ref{eq:equipartition}). We do not know the reason. Figure \ref{fig:sizePlots}(b) shows the size of the cloud versus $\Delta_{00}$. The cloud is smallest when $\Delta_{00}=-0.75\Gamma$ and grows rapidly as $|\Delta_{00}|$ increases, reflecting the decrease in $\omega$. Figure \ref{fig:sizePlots}(c) shows that the size of the cloud decreases as the magnetic field gradient increases up to the maximum value used in this work, $dB/dz=A=41$~G/cm. From equations (\ref{eq:frequency}) and (\ref{eq:equipartition}) we expect the relation $\sigma_{0} = \zeta (dB/dz)^{-1/2}$. This model fits quite well to the data, as shown in figure \ref{fig:sizePlots}(c), and gives a best fit parameter of $\zeta = 9.7(1)$~mm(G/cm)$^{1/2}$.

\subsection{Loss rate}
\label{sec:lifetime}

\begin{figure}[t]
	\centering
	\includegraphics[scale=0.7]{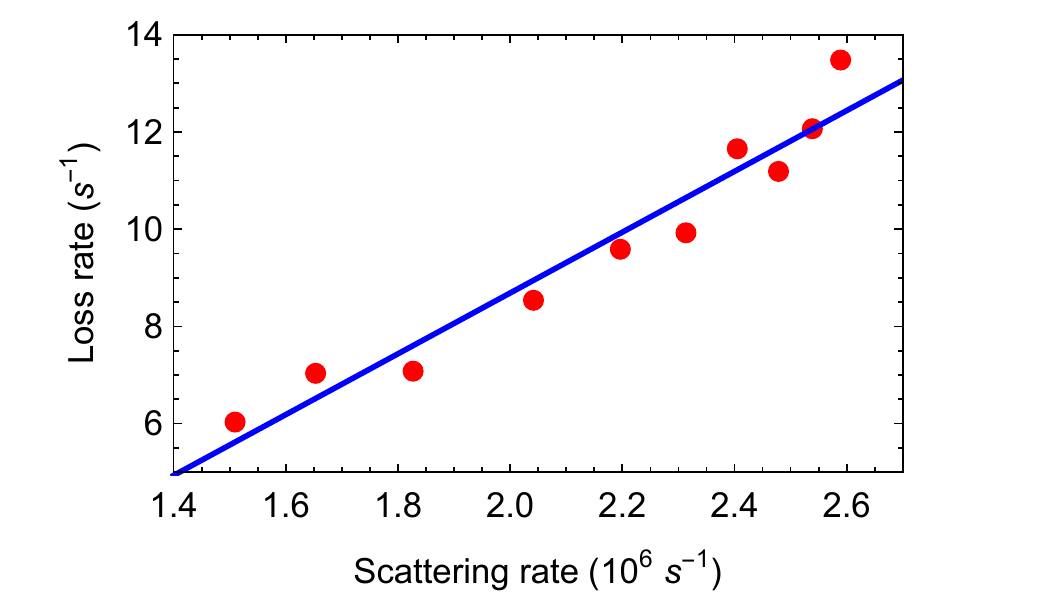}
	\caption{Loss rate versus scattering rate. Points: measurements (error bars are roughly the size of the points). Line: linear fit. }
	\label{fig:lossRate}
\end{figure}

For times $t > 50$~ms, the fluorescence decays exponentially as molecules are lost from the MOT. Figure \ref{fig:lossRate} shows the loss rate versus the scattering rate, which we control via the intensity $I_{00}$. The relation between $R_{\rm sc}$ and $I_{00}$ is obtained from the fit shown in figure \ref{fig:scatterPlots}(a). The loss rate increases approximately linearly with $R_{\rm sc}$ as we would expect if the loss is due to a leak out of the cooling cycle. The linear fit shown in figure \ref{fig:lossRate} gives a branching ratio for this leak of $6.3 (5) \times 10^{-6}$. The loss could be to a higher-lying vibrational state, or it could be due to magnetic dipole or electric quadrupole transitions that connect the excited state to rotational states $N=0$ and $N=2$. We note that the linear fit to the loss rate data gives a statistically significant negative intercept of $-3.8 (1.2)$~s$^{-1}$, which is not physical. This hints at a more complicated dependence on the scattering rate. 

When the temperature becomes comparable to the trap depth, the high energy molecules will spill out of the trap and increase the loss rate. This loss mechanism was considered in the context of the first molecular MOT~\cite{Barry2014}, where the rate for the process was approximated as 
\begin{equation}
R_{\rm loss} = \frac{2}{\pi} \omega_{\rho} e^{-m \omega_{\rho}^{2} r_{\rm trap}^{2}/(2 k_{\rm B} T)}.
\label{eq:evapLoss}
\end{equation} 
This result assumes that the oscillation is lightly damped and that the force is linear in the displacement out to the trap radius, $r_{\rm trap}$. These are poor approximations, but nevertheless we can expect the order of magnitude of the loss rate to be given by this equation. Our simulations of the MOT show that the restoring force has a turning point at a radial distance of about 8~mm. Choosing this value for $r_{\rm trap}$, and using our measured values of $\omega_{\rho}$ and $T$, equation (\ref{eq:evapLoss}) give $R_{\rm loss} = 1$~s$^{-1}$ at the highest intensity where the scattering rate is $2.6 \times 10^{6}$~s$^{-1}$. This contribution to the loss rate falls very rapidly as the intensity is reduced because $T$ falls faster than $\omega_{\rho}$. For example, lowering the scattering rate by 15\% reduces $R_{\rm loss}$ by a factor of 200. Thus, this loss mechanism could only be significant at the very highest intensity explored. Interestingly, the data point at the highest scattering rate in figure \ref{fig:lossRate} does indeed lie significantly above the linear fit to the data.

\subsection{Capture Velocity}
\label{sec:captureVelocity}

\begin{figure}[tb]
	\centering
	\includegraphics[scale=0.7]{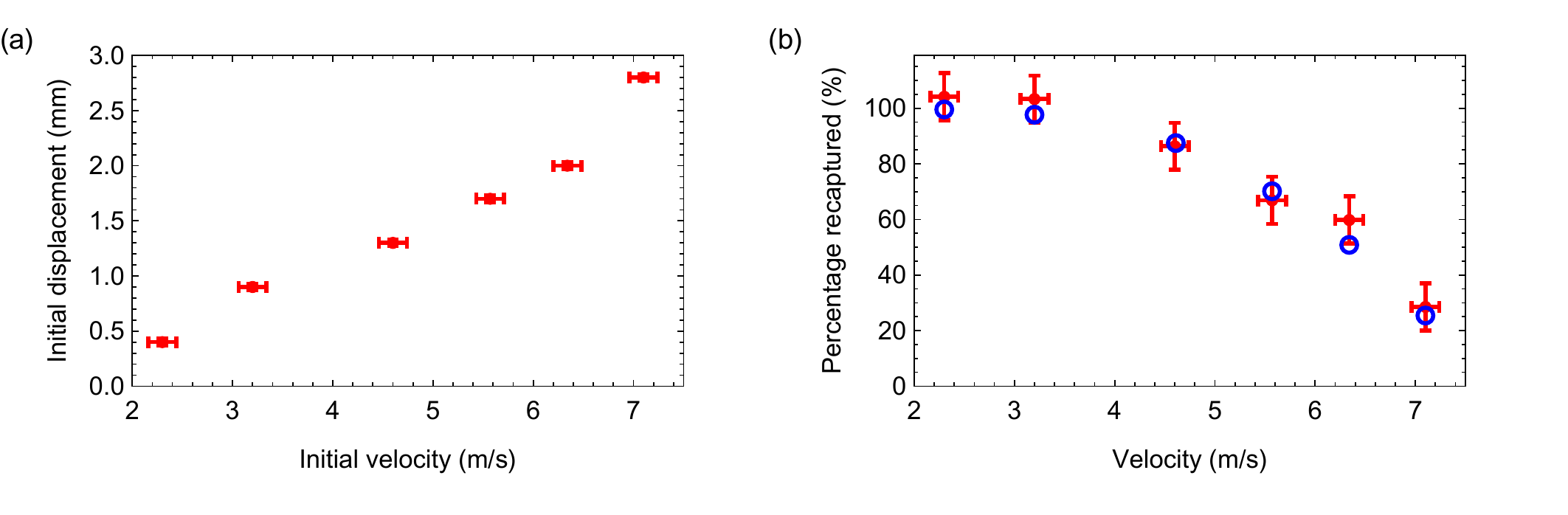}
	\caption{(a) Initial velocities and displacements used in measurements to determine the capture velocity. (b) Percentage of molecules recaptured into the MOT for each of the initial points in (a), indicated here by their velocities. Filled points: measurements. Open points: fit to the model described in the text.}
	\label{fig:escapeVelPlot}
\end{figure}

It is difficult to measure the capture velocity of the MOT directly. Instead, we measure the escape velocity by pushing the MOT and measuring the fraction of molecules that are lost as a function of their speed. We then infer the capture velocity from these results, with the help of a simple model. To apply an impulsive push, we turn off  ${\cal L}_{00}$ and pulse on the slowing light, ${\cal L}_{00}^{s}$, for a short time $t_{\rm push}$. The molecules are at the zero of the MOT magnetic field where states dark to the polarization of ${\cal L}_{00}^{s}$ are not destabilized effectively, so we found it necessary to modulate the polarization of ${\cal L}_{00}^{s}$ to reach a sufficient scattering rate. We also reduced the size of the slowing beam to $3$~mm $1/e^2$ radius, in order to increase the applied force. For various push parameters, we first determine the initial displacement and velocity of the cloud, $x_{\rm i}=x(t_{\rm push})$ and $v_{\rm i} = v(t_{\rm push})$, by imaging the cloud at various times after $t_{\rm push}$ with the MOT magnetic field turned off. Figure \ref{fig:escapeVelPlot}(a) shows the set of $\{x_{\rm i},v_{\rm i}\}$ pairs used. Then, for each push, we turn the MOT back on at $t=t_{\rm push}$ and measure the fraction of molecules recaptured by imaging the MOT at $t=t_{\rm push} + 20$~ms. Figure \ref{fig:escapeVelPlot}(b) shows this fraction versus $v_{\rm i}$, where the $v_{\rm i}$ should be understood as $\{x_{\rm i},v_{\rm i}\}$ pairs.

Were the initial displacements negligible, it would be simple to determine the capture velocity from these measurements. Since they are not, we use a model to interpret the results. In this model, the force in the direction ($x$) of the slowing light is given by equation (\ref{eq:totalForce}). To account for the intensity distribution of the MOT beams we make the replacement $s_{\rm eff}\rightarrow s_{\rm eff}(\frac{x}{\sqrt{2}})$ in equation (\ref{eq:force1DMOT}), where $s_{\rm eff}(r) = I_{00}(r)/I_{\rm s,eff}$, and $I_{00}(r)$ has a Gaussian intensity distribution with a $1/e^{2}$ radius of 8.1~mm, truncated at a radius of $r_{\rm trunc} = 15$~mm. Using this force, we solve the equation of motion for a distribution of initial coordinates in phase-space and for a time long enough that the distribution of final coordinates separates into two components, one with $x$ close to zero and the other with $x \gg r_{\rm trunc}$. This gives a contour in phase space, $v_{\rm sep}(x)$, which separates molecules that will be recaptured from those that escape. We then calculate the fraction of molecules from the initial phase-space distribution that lie inside this contour. We use a Gaussian spatial distribution centred at $x_{\rm i}$ with rms radius equal to the measured one, $\sigma_{0} = 2.0$~mm, and a Gaussian velocity distribution centred at $v_{\rm i}$ and characterized by the measured temperature of 12~mK. This gives the simulated recapture fraction for each $\{x_{\rm i},v_{\rm i}\}$ pair used for the measurement. In applying this procedure, it is not clear what values of $\Gamma_{\rm {eff}}$, $I_{\rm s, eff}$ and $g_{\rm eff}$ to use, so we keep them as free parameters. We compare the simulated results to the measured ones for a wide range of these free parameters and choose the parameter set that gives the smallest value of $\chi^{2}$. The open circles in figure \ref{fig:escapeVelPlot}(b) show the results that fit best. They are found for $\Gamma_{\rm eff} = 0.15 \Gamma$, $I_{\rm s,eff} = 50$~mW/cm$^{2}$, and $g_{\rm eff} = 0.25$, all reasonable values. This comparison between model and measurements gives us a best estimate for $v_{\rm sep}(x)$. The capture velocity, $v_{\rm c}$, is the largest velocity a molecule that starts at the edge of the MOT can have if it is to be recaptured, $v_{\rm c} = v_{\rm sep}(-r_{\rm trunc})$. The result is $v_{\rm c} = 11.2\pm^{1.2}_{2.0}$~m/s. The MOT simulations described in \cite{Tarbutt2015b} predicted a capture velocity of 20~m/s, but this used larger beams, higher power and a slightly different polarization configuration. Repeating these simulations for the exact parameters used in the experiment we find a capture velocity of 14~m/s, close to our measured value.

\section{Conclusions}

Despite the complexity of the level structure and the need to avoid optical pumping into dark states, the CaF MOT behaves much like a normal atomic MOT in many respects. The intensity dependence of the scattering rate, trap frequency and damping constant all conform to the analytical results based on an effective two-level model [equations (\ref{eq:Rsc}), (\ref{eq:frequency}) and (\ref{eq:damping})], and do so over a very wide range of intensities, although somewhat different values for the free parameters of these equations are needed in each case. The trap frequency is in excellent quantitative agreement with that predicted from rate equation simulations, whereas the scattering rate is a factor 2 smaller than expected and the damping constant is typically a factor of 5--10 smaller. We tentatively attribute the reduction of the damping constant to polarization gradient forces~\cite{Devlin2016}, though that remains to be verified. For our parameters, we measure a capture velocity of $11.2\pm^{1.2}_{2.0}$~m/s, consistent with our simulation. The temperature of the MOT is considerably higher than the Doppler temperature, especially at high intensity. It is also a factor of 2-3 higher than predicted by equation (\ref{eq:expectedTemperature}) when we use our measured values of the scattering rate and damping constant. This shows that the elevated temperature is a consequence of two factors, a reduced damping and some excess heating above that given by the diffusion constant in equation (\ref{eq:diffusionConstant}). The reasons for the reduced damping and the enhanced diffusion would make an interesting topic for future study. We find that the usual relation (equation (\ref{eq:equipartition})) between MOT size, temperature and trap frequency holds for our MOT, except at very low intensity where there seems to be a discrepancy. As expected, the size of the cloud scales inversely with the square root of the magnetic field gradient. Molecules are lost from the MOT with a typical rate of about 10~s$^{-1}$, and this loss rate scales linearly with scattering rate. The data is consistent with a leak out of the cooling cycle with a branching ratio of about $6\times 10^{-6}$. The lifetime, $\sim 100$~ms, is short by the standards of atomic MOTs, but is adequate for most purposes. Indeed, the whole process of capturing the molecules and cooling to sub-Doppler temperatures~\cite{Truppe2017b} takes less than 50~ms. Once loaded into a conservative trap, we can expect the lifetime to increase.

Rapid deceleration of the molecular beam with good velocity control is crucial to loading the largest number of molecules. It has also been one of the most difficult steps to perfect and understand. We use the frequency-chirp method to slow down the molecular beam. Our simulations and experiments~\cite{Truppe2017} suggest that this is a better slowing method than the alternative of broadening the frequency of the slowing laser~\cite{Barry2012}. We previously estimated a flux at the MOT region of $7\times 10^{5}$ molecules per cm$^{2}$ per shot with speeds below 15~m/s~\cite{Truppe2017}. We have recently re-evaluated the flux from our cryogenic source~\cite{Truppe2017c} and find that it is a factor of 4 smaller than our estimate in \cite{Truppe2017}. This reduces the flux of molecules with speeds below 15~m/s to $1.8\times 10^{5}$ molecules per cm$^{2}$ per shot. Scaling to our measured capture velocity, assuming the flux scales roughly as the square of the velocity, we expect about half this number. We use MOT beams with $1/e^{2}$ radius of 8.1~mm, giving a capture area of at least 0.5~cm$^{2}$. Thus, there are at least $4 \times 10^{4}$ trappable molecules, which is a factor of 2 larger than we observe in the MOT. The measured capture velocity is appropriate to molecules entering on axis and must be smaller away from the axis, and this may account for the discrepancy. While our simulations match well with our measured arrival-time distributions and velocity distributions, they predict a large peak in the number loaded into the MOT when the chirp parameters are tuned correctly, which we do not observe. Thus, while much progress has been made in understanding the slowing and MOT loading processes, some mysteries remain. We believe there is great scope for increasing the number of molecules loaded. We plan to add a region of transverse cooling before the slowing begins to compress the transverse velocity distribution, which should increase the number delivered to the MOT. Our linear frequency chirp is probably not the best chirp function and optimization of its functional form may give more molecules. Other slowing methods, such as Zeeman-Sisyphus deceleration~\cite{Fitch2016}, promise to deliver further increases in flux.  

At present, the only other 3D molecular MOTs reported are the dc and rf MOTs of SrF~\cite{Barry2014, McCarron2015, Norrgard2016, Steinecker2016} and the recent rf MOT of CaF~\cite{Anderegg2017}. For SrF, the rf MOT seems superior to the dc MOT. In particular, the rf MOT loads more molecules, has a long lifetime, and can be cooled towards the Doppler limit by lowering the laser intensity. By contrast, molecules in the dc MOT could not be cooled by lowering the intensity because its lifetime was found to decrease drastically at lower laser powers. Our dc MOT of CaF does not suffer from this problem. In fact, the lifetime is longer at lower intensities. The molecules cool as the intensity is lowered, just as in the SrF rf MOT. We suggest that the differences observed between the dc MOTs of CaF and SrF might be due to the difference in the ground-state hyperfine intervals. Most of the confinement in our CaF MOT comes from the dual-frequency effect~\cite{Tarbutt2015b} acting on the $F=2$ state. In CaF, the splitting between the upper two hyperfine components is about $3\Gamma$. When the detuning is $-\Gamma$, the upper $F=2$ level is addressed by a $\sigma^{-}$ component detuned by $-\Gamma$ and a $\sigma^{+}$ component detuned by about $2\Gamma$ (see figure \ref{fig:structure}(b)). It is known that this dual-frequency arrangement produces a strong confinement (see figure 2 of reference \cite{Tarbutt2015b}). The same polarization configuration has been used for the dc MOT of SrF~\cite{McCarron2015}, and the dual-frequency effect provides the confinement in that case too. In SrF however, the equivalent hyperfine splitting is $6\Gamma$, resulting in weaker confinement. It may be that at low laser intensity the confining forces in SrF are too weak, resulting in the short lifetimes observed. If this is the cause, then it could be solved by adding an extra frequency component to approach more closely the ideal dual-frequency scheme.

The molecule MOT presents many opportunities for new research. One is to study ultracold collisions between atoms and molecules and to cool molecules to even lower temperatures by sympathetic cooling with ultracold atoms~\cite{Lim2015}. Another is to load single molecules into optical tweezer traps in order to assemble small arrays~\cite{Barredo2016} for quantum simulation~\cite{Micheli2006}. The molecules could also be loaded into chip-scale electric traps where they could be coupled to a superconducting microwave resonator, realizing a molecular quantum processor~\cite{Andre2006}. Precise measurements of the vibrational frequency of CaF can test whether the fundamental constants are changing in time~\cite{Kajita2014}. The extension of the cooling and trapping methods to other amenable molecules~\cite{Smallman2014, Norrgard2017} can improve the measurements of the electric dipole moments of electron and proton~\cite{Tarbutt2013, Hunter2012} and advance the measurement of nuclear anapole moments~\cite{Cahn2014}.

\ack
We are grateful to Jack Devlin for his assistance and insight, and to Jon Dyne, Giovanni Marinaro and Valerijus Gerulis for their technical assistance. The research has received funding from EPSRC under grants EP/I012044, EP/M027716 and EP/P01058X/1, and from the European Research Council under the European Union's Seventh Framework Programme (FP7/2007-2013) / ERC grant agreement 320789.

\section*{References}
\bibliographystyle{iopart-num}
\bibliography{library}

\end{document}